\documentclass[lettersize,journal]{IEEEtran}
\usepackage{amsmath,amsfonts}
\usepackage{algorithmic}
\usepackage{algorithm}
\usepackage{array}
\usepackage[caption=false,font=normalsize,labelfont=sf,textfont=sf]{subfig}
\usepackage{textcomp}
\usepackage{stfloats}
\usepackage{url}
\usepackage{verbatim}
\usepackage{graphicx}
\usepackage{cite}
\begin{document}

\title{HarmoGAN: Harmony-Aware Music-driven Motion Synthesis with Perceptual Constraint on UGC Datasets}

\author{Xinyi~Wu,
        Haohong~Wang,~\IEEEmembership{Member,~IEEE},
        and~Aggelos~K. Katsaggelos,~\IEEEmembership{Fellow,~IEEE}
\thanks{This work was supported by TCL Research America through Interactive Hyperstory project.}
\thanks{X. Wu and A. K. Katsaggelos are with the Department of Computer Science and Electrical Engineering, Northwestern University, Evanston, IL 60208 USA (e-mail:xinyiwu2019.1@u.northwestern.edu;a-katsaggelos@northwestern.edu).}
\thanks{H. Wang is with TCL Research America, San Jose, CA 95110 USA (e-mail: haohong.wang@tcl.com).}
\thanks{Manuscript received April 19, 2021; revised August 16, 2021.}}

\markboth{IEEE TRANSACTIONS ON CIRCUITS AND SYSTEMS FOR VIDEO TECHNOLOGY,~Vol.~14, No.~8, August~2021}%
{Wu \MakeLowercase{\textit{et al.}}: Harmony-aware Human Motion Synthesis with Music}

\IEEEpubid{0000--0000/00\$00.00~\copyright~2021 IEEE}

\maketitle

\begin{abstract}
With the popularity of video-based user-generated content (UGC) on social media, harmony, as dictated by human perceptual principles, is critical in assessing the rhythmic consistency of audio-visual UGCs for better user engagement. In this work, we propose a novel harmony-aware GAN framework, following a specifically designed harmony evaluation strategy to enhance rhythmic synchronization in the automatic music-to-motion synthesis using a UGC dance dataset. This harmony strategy utilizes refined cross-modal beat detection to capture closely correlated audio and visual rhythms in an audio-visual pair. To mimic human attention mechanism, we introduce saliency-based beat weighting and interval-driven beat alignment, which ensures accurate harmony score estimation consistent with human perception. Building on this strategy, our model, employing efficient encoder-decoder and depth-lifting designs, is adversarially trained based on categorized musical meter segments to generate realistic and rhythmic 3D human motions. We further incorporate our harmony evaluation strategy as a weakly supervised perceptual constraint to flexibly guide the synchronized audio-visual rhythms during the generation process.  Experimental results show that our proposed model significantly outperforms other leading music-to-motion methods in rhythmic harmony, both quantitatively and qualitatively, even with limited UGC training data. Live samples 15
can be watched at: https://youtu.be/tWwz7yq4aUs.

\end{abstract}


\section{Introduction}

\IEEEPARstart{T}{he} development of various social media platforms has facilitated the popularity of audio-visual UGC videos as an essential part of daily entertainment. To produce engaging high-quality content, rhythmic harmony is crucial for providing viewers with perceptual immersion and pleasure \cite{davis2018visual}. However, due to excessive demands of human labor and professional knowledge, creators often struggle to continuously design appealing audio-visual creations, particularly visual sequences, with synchronized cross-modal rhythms.

\begin{figure}[htbp]
\centerline{\includegraphics[width=0.45\textwidth]{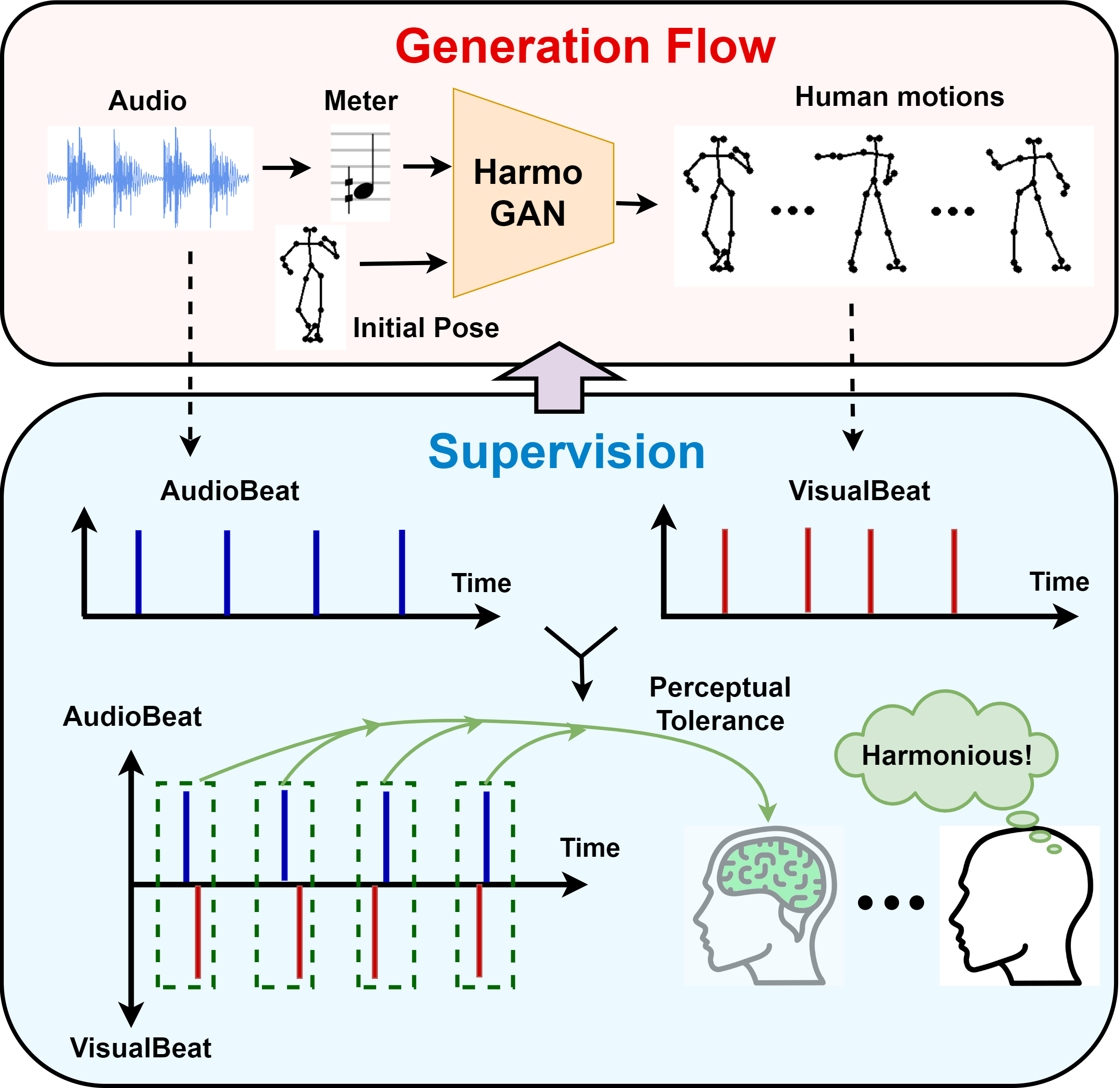}}
\caption{Illustration of the overall workflow. The proposed generation flow handles the music-to-motion synthesis based on musical meters. We incorporate a novel regularizer following our designed harmony evaluation strategy to supervise the motion generation through refined audio-visual beat alignment at the perceptual level.} 
\end{figure}

Music-driven motion synthesis has recently emerged as a solution to this issue. It enables creators to flexibly generate realistic human motions corresponding to a given music piece by analyzing the audio-visual mapping \cite{sun2020deepdance}. As a cross-modal generation task, additional strategies, such as introducing priors \cite{tang2018dance, wang2022groupdancer, kim2022brand} and regularizations \cite{yalta2019weakly, duan2021automatic, Aristidou2022dancer}, are required to control the generation quality. This involves not only the spatio-temporal realism of the generated visual sequence but also the stylistic consistency and rhythmic synchronization between the audio-visual pair.

\IEEEpubidadjcol

Traditional methods address the music-to-motion task by utilizing an audio-visual database. They extract cross-modal features in analytically and compare feature-level similarity to retrieve the best-matching visual sequence based on the input music \cite{shiratori2006dancing, lee2013music}. Constrained by the richness of the resource database, such methods lack flexibility for the synthesized motions. Learning-based approaches have gradually come to dominate the field. Early methods like Hidden Markov Models (HMMs) \cite{ofli2010multi, ofli2011learn2dance} or graph-based optimization \cite{fan2011example}, are proved effective for continuous and flexible motion generation. Yet, their limited ability to capture contextual information can cause difficulty for producing realistic and harmonious audio-visual results.

As deep neural networks demonstrate their powerful ability to handle generative tasks, they also enable learning the end-to-end mapping between the audio and visual sequences. The success of \cite{lee2018listen} shows that the encoder-decoder network can efficiently translate the auditory features into choreographic poses. Following this trace, architectures such as Long short-term memory (LSTM) \cite{tang2018dance}, Gated recurrent unit (GRU) \cite{wang2022groupdancer,hou2023graph}, and Recurrent Neural Network (RNN) \cite{wang2019combining, ahn2020generative} present advancements on strengthening temporal smoothness of the generated human motions. To further improve the stylistic consistency during the audio-visual feature transformation, \cite{ren2020self,sun2020deepdance} leverage Generative Adversarial Networks (GANs), which additionally include discriminator to adversarially supervise the similarity between cross-modal features. Recent attention-driven Transformers \cite{kim2022brand,le2023music} or diffusion-based models \cite{tseng2023edge} also significantly boost the naturalness of music-to-motion synthesis but heavily rely on resource-intensive platforms and lack micro-level rhythm controllability.

Practical experience has proved that rhythm-space harmony greatly impacts the perceptual quality of generated audio-visual creations \cite{chu2011rhythm,ho2013extraction}. Most Existing methods address this issue limitedly by either introducing additional music rhythm features as prior \cite{gao2022music, wang2022groupdancer, Aristidou2022dancer} or rely on implicitly refining rhythm learning in the hidden space \cite{li2021ai,valle2021transflower,li2022danceformer}. Research in the video generation area indicates that, beats extracted from both audio and visual sequences are highly correlated with the perceptual cross-modal rhythmic synchronization \cite{davis2018visual,bellini2018dance}. Leveraging this insight, \cite{lee2019dancing,au2022choreograph} apply a post-processing warper to force the generated human motions synchronized to audio beats obtained from music input. While this enhances the audio-visual harmony to some extent, it can damage the coherence of synthesized motions. To achieve explicit regularization, \cite{yalta2019weakly} propose a beat-matching loss function to flexibly align audio and visual beats in the resulting audio-visual pair. However, its coarse beat extraction strategy and lack of considering human perceptual principles constrain the potential perceptual improvement. Therefore, there remains a need for a method that can cross-modally generate realistic motion sequences while preserving fine-detailed rhythmic harmony consistent with human perception.  

In this paper, we propose a harmony-aware GAN-based model to efficiently tackle music-driven motion synthesis with precisely controlled cross-modal rhythmic consistency at beat and meter levels. To achieve this, we design a harmony evaluation strategy building on beat alignment approaches from \cite{davis2018visual,bellini2018dance,yalta2019weakly}. By introducing human perceptual principles, this strategy refines audio-visual beat extraction, particularly for the visual sequence, with saliency-based weighting to mimic the human attention mechanism. We then analyze cross-modal beat intervals to derive a score for rhythmic harmony that is consistent with human perception. To cooperate with such a harmony strategy, we further utilize
an adversarial training framework driven by music meter segments, which provide high-level rhythmic priors based on categorized beat groups. This not only enriches the understanding of music rhythms but also implicitly guides the coherence and smoothness of generated motions both within and across segments. Our framework consists of an encoder-decoder generator and two discriminators to ensure cross-modal stylistic consistency and spatio-temporal motion realism. We additionally integrate a depth-lifting network, which facilitates realistic 3D pose generation with less learning pressure via efficient 2D-to-3D pose lifting. To strengthen the learning of rhythmic synchronization, we also incorporate our harmony evaluation strategy as a weakly supervised perceptual constraint to flexibly control the visual rhythms in our generation. This enables us to enhance generalizability even trained on a customized UGC dataset with relatively fewer samples. Experimental results demonstrate that our model outperforms the other leading music-to-motion methods in terms of optimally balancing motion realism and cross-modal rhythmic harmony.

The contributions of our work can be summarized as follows:
\vspace{1mm}
\begin{itemize}
   
\item We propose a novel harmony-aware music-to-motion synthesis framework driven by the analysis of high-level perceptual rhythm representations involving musical meters and beats. This allows us to effectively balance motion realism and audio-visual rhythmic synchronization.
    
\item We design an efficient encoder-decoder GAN with dual discriminators and an additional depth-lifting network. This alleviates learning pressure via the 2D-to-3D pose lifting and comprehensively ensures natural 3D human motion generation with adversarially controlled cross-modal styles and spatio-temporal pose flows.  

\item We introduce a refined rhythmic harmony evaluation strategy considering human perceptual principles and integrate it as a weakly supervised constraint. This flexibly regularizes audio-visual rhythm synchronization with better generalizability on limited UGC training data. Experimental results show that our model effectively aligns audio and visual rhythms at the perceptual level, outperforming other leading methods.

\end{itemize}

\section{Related Work}

In this section, we first review the early-stage progress of research on music-driven motion synthesis, which emphasizes two key improvement aspects: spatio-temporal motion realism and cross-modal style consistency. As harmony gradually plays a crucial role in audio-visual generative tasks, we also analyze potential critical factors and existing feasible approaches that contribute to perceptual rhythmic harmony.

\subsection{Facilitating Music-driven Motion Synthesis}

\textbf{Spatio-temporal motion realism:} For smooth and continuous motion synthesis, traditional methods tackled music-to-motion tasks by retrieving the best-matching candidate in the motion database. This is done through calculating the feature similarity given an audio-visual pair using algorithms such as histogram \cite{shiratori2006dancing} or clustering \cite{lee2013music}. To flexibly enrich the generated motion sequences, Hidden Markov Model (HMM) \cite{ofli2010multi,ofli2011learn2dance} and optimization methods (e.g. Graph \cite{fan2011example}) have proven effective for producing diverse motion poses based on the extracted fine-grained musical feature and contextual motion information. 

To further improve the spatio-temporal coherence of the generated motion, deep neural networks have dominated the field by effectively capturing deep features to estimate an end-to-end mapping between audio and visual sequences. \cite{lee2018listen} first trained an efficient network consisting of a music encoder and a motion decoder to facilitate the audio-to-visual transformation. As temporal coherence is a critical and complex component influencing overall motion generation quality, building on previous deep methods,  \cite{tang2018dance} claimed that the use of long short-term memory (LSTM) blocks could benefit the extraction of temporal auditory features with memorized hidden states and yield smooth 3D body poses. 

Moreover, \cite{wang2019combining} utilized Recurrent Neural Networks (RNN) while \cite{ahn2020generative} leveraged Autoregressive Neural Networks (ARNN), which have been found powerful for realistic long-term generation by thoroughly exploiting contextual features, especially in the temporal domain. Further advancements such as utilizing curriculum learning strategy \cite{huang2021DR},  or introducing additional constraints like coarse-to-fine pose hierarchy loss \cite{Aristidou2022dancer} or foot-sliding contact loss \cite{tseng2023edge}, have also effectively enhanced motion realism to the next level. Even though these methods are noted for improving motion realism, as a cross-modal task, there is a need to manage the stylistic coherence between the given music piece and the synthesized human motions for satisfying user requirements. 

\textbf{Cross-modal style consistency:} As Generative Adversarial Networks (GANs) demonstrated their effectiveness for generative tasks, \cite{lee2019dancing} introduced an adversarial framework to regularize the generator with an additional discriminator architecture that supervises the cross-modal consistency of the audio-visual pair. \cite{ren2020self} made further advancements by employing a self-attention global content discriminator to refine features extracted by audio-visual branches during the supervision. \cite{gao2022music} utilized a multi-feature fusion strategy to capture the stylistic features of music sequences as priors during the cross-modal feature transformation.  

To strengthen the control of the generated motion style, \cite{sun2020deepdance} trained a genre-specific dance network to obtain fluent and consistent choreographic movements in different genres. \cite{ferreira2021learning} introduced a dense music style vector representation, which works as a categorical dictionary corresponding to dance style classes in high-dimensional space. \cite{starke2022deepphase} similarly integrated musical features into phase manifolds for effectively synthesizing motions with desired styles. Moreover, \cite{hou2023graph} regularized the audio-visual style consistency by a reverse-generation loss, allowing generated motions to transfer back to acoustic features in alignment with the given music. \cite{wang2022groupdancer} then extended the single-person scenario to multi-person, achieving motion-style collaboration under mixed group dancer training and selective updating.

Transformer architectures have then demonstrated their power in sequence-to-sequence tasks. \cite{valle2021transflower} proposed a Full Attention Cross-modal Transformer (FACT) to effectively handle audio-to-visual transformation using multi-head dense attention. Building on this, \cite{li2021ai} leveraged additional transforms to better encode stylistic features from music and initial poses. Moving forward, \cite{kim2022brand} combined adversarial learning with transformers to create a scalable model for multi-genre dance generation, while \cite{siyao2022bailando} designed an actor-critic Generative Pre-trained Transformer (GPT) to produce choreographic memory units coherent with music. \cite{le2023music} further advanced this by expanding the single-dancer generation to multi-style group dancing using only a music sequence input. Despite these enhancements in cross-modal style consistency, most methods require large training datasets and lack precise control over rhythmic synchronization between audio-visual pairs, which significantly impacts viewer perception.

\subsection{Exploring Audio-visual Rhythmic Harmony}

The rise of UGC videos on social media drives creators to produce pleasant audio-visual content that deeply immerses viewers. Given harmony is a subjective perception, research reveals that audio-visual harmony involves several perceptual aspects, such as spatial localization \cite{xuan2021discriminative}, temporal alignment \cite{Iashin_2022_BMVC}, or semantic consistency \cite{lee2022audio}. Among these, the synchronization between audio-visual rhythms is one of the most crucial factors for creating harmonious and engaging creations.

In early literature, rhythmic synchronization was typically achieved by introducing additional rhythm features as priors, including MFCC \cite{huang2021DR, li2021ai}, tempo \cite{chen2021choreomaster, Aristidou2022dancer}, and onset strength \cite{tang2018dance, valle2021transflower}. Further advancements relied on cross-modal network architectures such as GANs \cite{ren2020self,gao2022music} and Transformers \cite{li2021ai,li2022danceformer} to implicitly transfer learned audio rhythms into synchronized visual rhythms during generation. Additionally, \cite{duan2021automatic} decoded music embeddings back to rhythm features and conducted alignment to enhance precise audio rhythms captured in the encoding phase. However, these methods lack explicit definitions and corresponding regularizations to ensure rhythmic synchronization.

According to musicology, the concepts of meter and beat provide clear and precise representations of musical rhythm \cite{benward2014music}. Building on this, \cite{chu2011rhythm} pioneered rhythm-based cross-media alignment for dance videos by extracting beats from significant signal changes and matching audio-visual segments to ensure synchronized cross-modal rhythms. \cite{davis2018visual} enhanced this approach using the onset envelope for more precise beat extraction and employed a curve-based time warping function for efficient rhythmic alignment. \cite{lee2019dancing} further utilized a post-processing warper to align extracted audio beats with visual beats from the generated motion. To avoid unnatural motion due to forced warping, \cite{au2022choreograph} introduced dual warpers to handle cross-modal rhythmic synchronization while preserving original motion flows. Unlike multi-stage methods, \cite{bellini2018dance} integrated frame-level direction-based beat extraction and alignment into a compatibility score for dynamic synchronization. For real-time adjustments of realistic rhythmic motions, \cite{yalta2019weakly} designed a novel loss function to guide rhythmic alignment during motion synthesis using entropic distance between derived visual beats and musical features.

Rhythmic harmony, as a subjective experience, is significantly influenced by human perceptual principles. \cite{ho2013extraction} discovered that human reaction times exist a delay, thus cross-modal rhythm alignment could be relaxed to better match perceptual principles. Additionally, attention mechanisms (e.g., inattentional blindness \cite{mack1998inattentional}), which also greatly influence the perception of auditory and visual beats, should be incorporated into beat extraction and alignment to accurately imitate human perception. Yet, few audio-to-visual methods comprehensively address these perceptual factors. There remains a need for a deep learning method that enables refined rhythmic analysis at both meter and beat levels, aligns with human perceptual principles, and provides flexible, dynamic regularization of synthesized visual sequences to enhance viewer engagement.

\section{Exploring audio-visual harmony}\label{AVH-Sec3}

This section discusses our designed harmony evaluation strategy for assessing rhythmic synchronization in audio-to-visual generation tasks. According to human perceptual principles,  we conduct a refined beat extraction and attention-based beat alignment to derive a more precise cross-modal harmony score consistent with viewer perception.

\subsection{From Harmony to Beat}

Harmony plays an essential role in evaluating the quality of audio-visual creations, where rhythmic synchronization is particularly emphasized for benefiting immersion. Since the senses of vision and hearing are closely related and mutually influential in brain processing, given an auditory sequence $A=\{a_1, a_2,..., a_t\}$ and a visual sequence $V=\{v_1, v_2,..., v_t\}$, their rhythms are supposed to be temporally consistent to achieve rhythmic harmony.

Based on musicology, musical rhythms can be effectively represented by musical meter \cite{benward2014music}, which is further concretized as a sequence of audio beats $B_a=\{ba_1, ba_2,..., ba_N\}$, denoting time positions at which the auditory amplitude is distinctly changed in the signal \cite{scheirer1998tempo}. Correspondingly, visual beats $B_v=\{bv_1, bv_2,..., bv_M\}$ have also been proposed for visual media to depict visual rhythms \cite{chu2011rhythm,davis2018visual}. The audio-visual rhythmic harmony can thus be defined as the synchronization between every audio beat and visual beat throughout the cross-modal sequence pair \cite{bellini2018dance, yalta2019weakly}.

\begin{figure}[htbp]
\centerline{\includegraphics[scale=0.35]{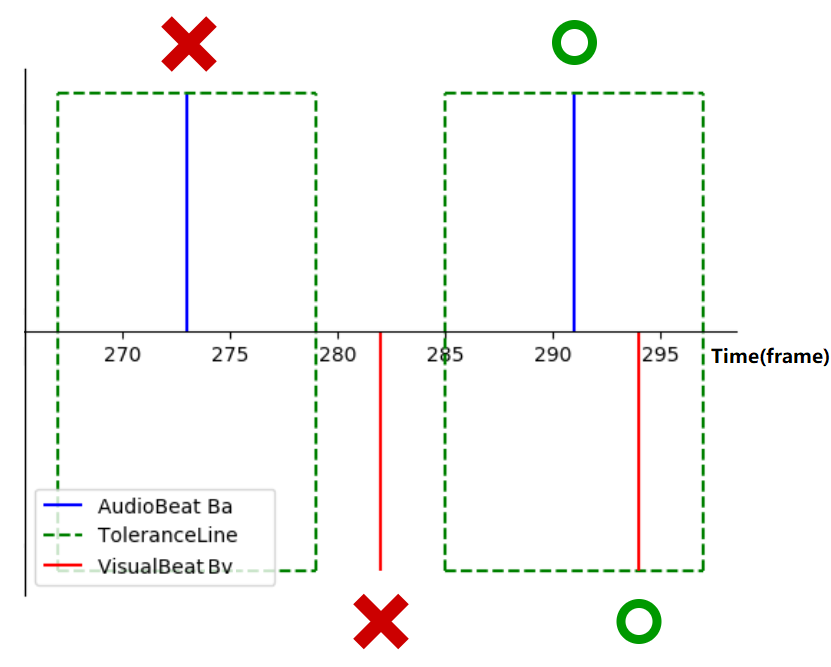}}
\caption{Example of synchronized and unsynchronized audio-visual beat pairs, labeled by green circles and red crosses, respectively, under tolerance fields neighboring each audio beat. For clear visualization, we assign 1 in y axis for $B_a$ while -1 for $B_v$.}
\label{BeatDismatch}
\end{figure}

As findings in \cite{ho2013extraction} have confirmed there exists a reaction delay when observing audio-visual creations, we introduce tolerance fields to better imitate the perceptual judgment of the synchronization status between a pair of cross-modal beats, which are illustrated in Fig. \ref{BeatDismatch}. Following such ideas, the rhythmic harmony $h$ can be quantified by aligning the extracted audio and visual beats as:
\begin{equation}\label{basic}
h=L(F_a(A),F_v(V))=L(B_a,B_v),
\end{equation}
where $F_a$ and $F_v$ denote beat detection methods conducted on the audio and visual signals, respectively, while $L$ represents the alignment algorithm.

Due to the increased interest in research on human-related videos, we aim to analyze an evaluation strategy that precisely captures perceptual rhythmic synchronization between music and human motions. This comprises refined consistent cross-modal beat extraction and attention-driven beat alignment considering human perceptual principles.  

\subsection{Refined Audio-visual Beat Extraction}\label{sec3.2}

To detect audio beats $B_a$, mainstream approaches (e.g. \cite{lee2019dancing, li2021ai,wang2022groupdancer}) utilize Mel spectrogram to obtain onset strengths as auditory amplitude and extract time points of local maximums as the occurrence of beats. Yet, in terms of visual beats  $B_v$, strategies are varied based on frame-level or pose-level visual media. The frame-based visual beat detection leverages optical flow estimation to represent the visual amplitude \cite{davis2018visual,bellini2018dance}, while these methods are sensitive to background movements. Focusing on the foreground human motion, methods driven by extracted body poses are more robust and feasible to process rhythmic audio-visual human video \cite{lee2019dancing,ren2020self}. To pursue detecting visual beats that are consistent with ground-truth musical rhythm features, \cite{yalta2019weakly} have proposed a state-of-the-art approach using joint-wise standard deviation (SD) to derive directional motion changes as beats.

\begin{figure}[htbp]
\centerline{\includegraphics[scale=0.13]{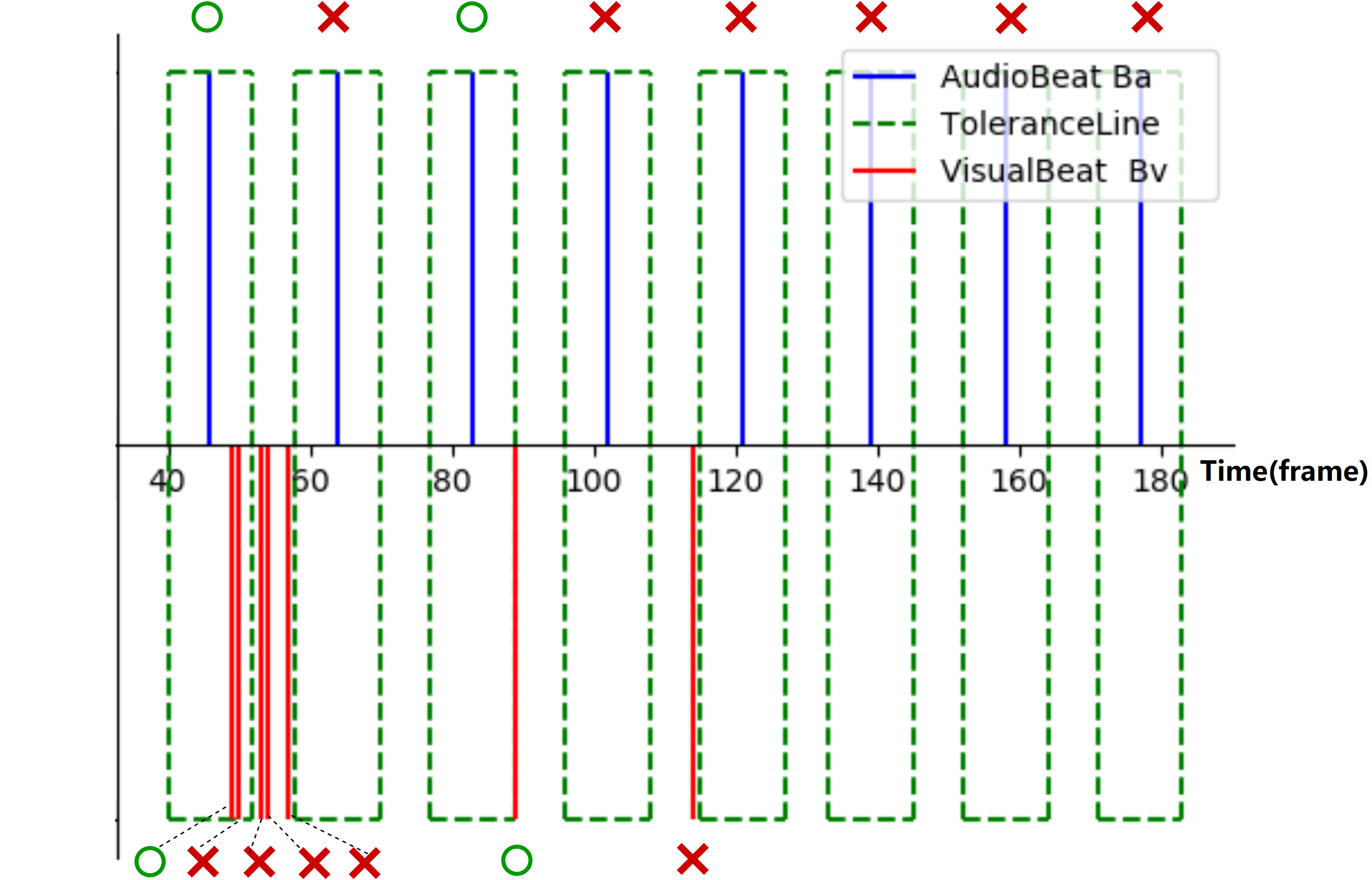}}
\caption{Distortion of rhythmic harmony between onset-based audio beats and SD-driven visual beats. Unsynchronized beat pairs are frequently present due to omissions or redundancies.} 
\label{AVcmp1}
\end{figure}

However, such a SD-based visual beat detection depicts visual rhythms at a coarse level. When performed on the Dance dataset \cite{tang2018dance} with professional audio-visual dancing sequences, in Fig. \ref{AVcmp1} it can be observed that SD-based visual beats are rarely consistent with the mainstream onset-driven audio beats. Speed, as one of the important factors to represent trends of human motions in cross-modal tasks \cite{ho2013extraction, lee2019dancing}, is not effectively considered in the motion SD approach \cite{yalta2019weakly}. Therefore, we refine the extraction of visual beats by introducing a fine-detailed analysis of joint velocity to improve the derived audio-visual beat consistency. 

In \cite{ye2020choreonet, Aristidou2022dancer}, dance-like human motions have been proven to be composed of several movement units (e.g., hand lift) that are highly consistent and correlated with musical rhythms. Hence, we manage to divide movement units across the entire motion sequence $V$ for defining visual beats well-aligned with audio beats. This is done by analyzing the joint velocity sum $J=\{j_{2}, j_{3},...,j_{t}\}$, which represents visual rhythms at the kinematic level, as: 

\begin{equation}
j_{t}=\sum_{p=1}^{P}{|v_{t}^{p}-v_{t-1}^{p}|},
\end{equation}
where $v_{t}^{p}$ denotes the location of $p^{th}$ joint in the body pose at time $t$ while $P$ is the total number of joints.

\begin{figure}[htbp]
\centerline{\includegraphics[scale=0.06]{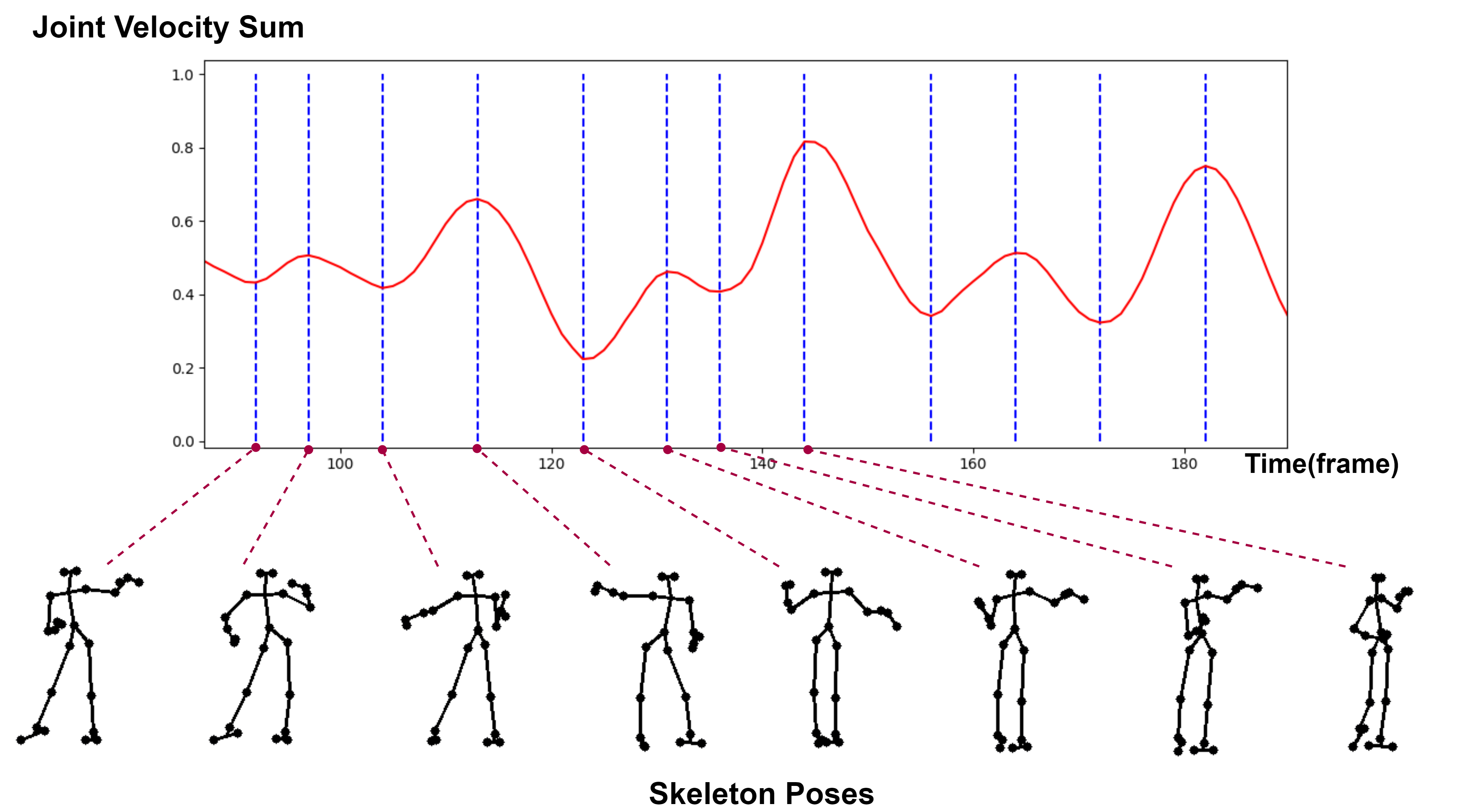}}
\caption{Relationship between joint velocity sum $J$ and evolution of human motions. The changes of movement units generally accompany the local minimum and maximum in $J$.}
\label{Jvt-T}
\end{figure}

As demonstrated in Fig. \ref{Jvt-T}, the joint velocity sum $J$ empirically shows a great potential to get correlated with movement units. When $J$ approaches its local maximum or minimum, we can observe that a single movement unit is completed based on human perception. Therefore, our visual beats $B_{v}$ can be defined by collecting time positions of local maximum and minimum in $J$ as:

\begin{equation}\label{vsb}
\begin{aligned}
    \Delta j_t &= j_t - j_{t-1}, \\
    B_v &= \{ t \mid (\Delta j_{t-1} > 0 \text{ and } \Delta j_t < 0) \\
    & \quad \quad \text{ or } (\Delta j_{t-1} < 0 \text{ and } \Delta j_t > 0) \}
\end{aligned}
\end{equation}

\begin{figure}[htbp]
\centerline{\includegraphics[scale=0.13]{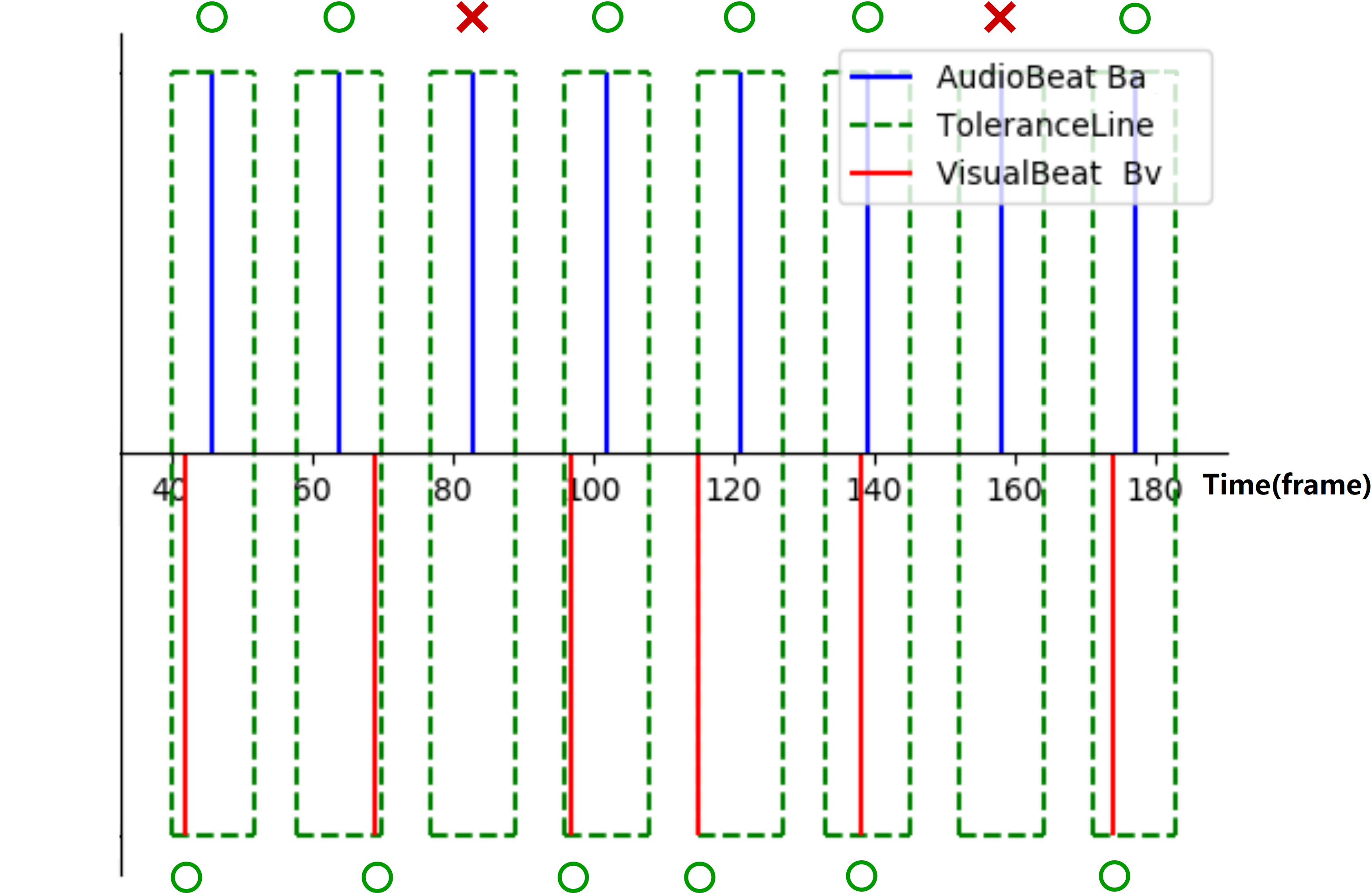}}
\caption{Improved beat synchronization between our extracted visual beats and onset-driven audio beats tested under the same conditions with Fig. \ref{AVcmp1}.} 
\label{AVcmp2}
\end{figure}

Compared to SD-based visual beat detection shown in Fig. \ref{AVcmp1}, our method, demonstrated in Fig. \ref{AVcmp2}, makes advancements by effectively reducing omitted and redundant audio-visual beat pairs for enhanced rhythmic synchronization over ground-truth professional dancer. Note that lacking the consideration of onset-based audio beats in \cite{yalta2019weakly} can also partially explain its degraded performance. 

In addition to refining beat extraction, attention, which is a crucial perceptual factor, is also integrated into our subsequent beat alignment method to appropriately quantify audio-visual rhythmic harmony that precisely imitates human perception.

\subsection{Attention-based Beat Alignment}\label{avh-attention}

\textbf{Saliency weighting:} Psychophysical studies reveal that non-salient objects can be perceptually neglected without any awareness following attention mechanisms \cite{mack1998inattentional}. Such cases influence not only vision but also hearing systems by causing unintentional blindness and deafness \cite{dehais2014failure}. We thus introduce the concept of saliency-based weighting \cite{dai2019adaptive,zhu2018saliency} to preserve only high-saliency audio and visual beats that are observable during subjective judgment of rhythmic harmony.

According to subsection \ref{sec3.2}, we collect the corresponding amplitude values, represented by onset strength and joint velocity sum $J$ for $B_a$ and $B_v$, as audio beat saliency  $S_a=\{sa_{1},sa_{2},...,sa_{N}\}$ and visual beat saliency $S_v=\{sv_{1},sv_{2},...,sv_{M}\}$, respectively.   

Inspired by the SD approach in \cite{yalta2019weakly}, which efficiently detects salient events, we construct adaptive weighting masks using global SD of beat saliency. The auditory mask $W_a=\{wa_{1},wa_{2},...,wa_{N}\}$ is formed as:

\begin{equation}\label{filta}
wa_{i}=
\begin{cases}
    1 & \text{if $sa_{i}-\lambda_{1}\sigma_{S_a}>0$} \\
    0 & \text{otherwise,}
  \end{cases}
\end{equation}
where $i$ iterates sequence elements and $\sigma_{S_a}$ represents the standard deviation of $S_a$. $\lambda_{1}$ is a constant scale factor for adjusting the audio saliency threshold.

Differently from only detecting local maximum for audio beats, our visual beats are extracted by collecting both local maximum and minimum of $J$ to better depict motion rhythms. Avoiding that our SD-based weighting strategy always penalizes the beats from local minimum, we substitute the original $S_v$ with $S^{*}_v$ by maximum-minimum differencing as:

\begin{equation}
sv_{i}^{*}=
\begin{cases}
   |sv_{i}-sv_{i-1}| &
   \text{$i=2,...,M$}, \\
   |sv_{i}- j| & \text{$i=1$},
 \end{cases}
\end{equation}
where $j$ denotes the initial value of $J$.

The visual mask $W_v=\{wv_{1},wv_{2},...,wv_{M}\}$ can then be formed by:
\begin{equation}\label{filtv}
wv_{i}=
\begin{cases}
    1 & \text{if $sv_{i}^{*}-\lambda_{2}\sigma_{S^{*}_v}>0$} \\
    0 & \text{otherwise,}
  \end{cases}
\end{equation}
where $\sigma_{S^{*}_v}$ denotes the standard deviation of $S^{*}_v$ and $\lambda_{2}$ is a constant adjusting the visual saliency threshold. 

By integrating saliency masks $W_a$ and $W_v$ using element-wise multiplication with our beat-based rhythmic representations, we can obtain $N'~ (N' < N)$ salient audio beats $B’_a$ and $M’~ (M’ < M)$ salient visual beats $B’_{v}$ as well as their corresponding saliency values $S’_a$ and  $S’_{v}$, respectively, as:

\begin{equation}
\begin{aligned}
B’_a= \{ ba_{i} \mid wa_{i} \cdot ba_{i} > 0, \; i = 1, 2, \ldots, N \} \\
S’_a= \{ sa_{i} \mid wa_{i} \cdot sa_{i} > 0, \; i = 1, 2, \ldots, N \}
 \end{aligned}
\end{equation}

\begin{equation}
\begin{aligned}
B’_v= \{ bv_{i} \mid wv_{i} \cdot bv_{i} > 0, \; i = 1, 2, \ldots, M \} \\
S’_v= \{ sv_{i}^{*} \mid wv_{i} \cdot sv_{i}^{*} > 0, \; i = 1, 2, \ldots, M \}
 \end{aligned}
\end{equation}

This allows us to mimic the impacts of human attention on capturing cross-modal beats practically and enhances the accuracy of our derived harmony score in the next phase.   

\vspace{1mm}
\textbf{Interval-driven beat alignment: } Given that harmonious feelings in audio-visual human perception are influenced by fuzzy brain processing and biological delays \cite{bay2003survey,picot2011line}, we designed an interval-driven beat alignment method to better simulate such harmony perception. During this process, we derive a score $h$, which quantifies the amplitude of cross-modal harmony in an arbitrary audio-visual pair.

Instead of handling continuous features like in \cite{duan2021automatic,gao2022music}, we adopt the strategy illustrated in Fig. \ref{BeatDismatch}, where we align perceptually salient audio and visual beats in a discrete manner. These beats, extracted to effectively represent the rhythms of the given music and motions, are compared based on their cross-modal time intervals and a pre-defined human perception delay \cite{ho2013extraction}. This leads us to determine whether each audio-visual beat pair is synchronized or not. By counting the overall synchronization status, we can accurately obtain the harmony score $h$ that efficiently imitates brain perception.

In the previous steps, $N'$ salient audio beats $B’_a=\{ba'_1,ba’_2,...,ba’_{N’}\}$ and $M’$ salient visual beats $B’_{v}=\{bv’_1,bv’_2,...,bv’_{M’}\}$ have been derived. Anchored by each audio beat $ba'_i$, we are able to target the corresponding audio-visual beat pair by locating the nearest visual beat and determine its synchronization status $hp_i$ according to the cross-modal beat interval as:

\begin{equation}\label{Tdelay}
hp_i = \left\{
\begin{array}{ll}
1, & \text{if } \min\limits_{k} |ba'_i - bv'_k| < T_{\text{delay}} \\
0, & \text{otherwise}
\end{array}
\right., i = 1, 2, \ldots, N'
\end{equation}
where $i$ and $k$ iterate beats in $B’_a$ and $B’_{v}$, respectively. $T_{delay}$ is a constant perceptual time delay.  

That is to say, only audio-visual beat pairs within perceptual tolerance field can be considered as valid rhythmic synchronization that contributes to the sense of cross-modal harmony. We further introduce beat saliency $S’_a=\{sa'_1,sa’_2,...,sa’_{N’}\}$ to conduct a weighted sum for balancing the count of total synchronized beat pairs, $h_s$, as:

\begin{equation}\label{hitsum}
h_s=\sum_{i=1}^{N’} hp_i \times sa’_{i}.
\end{equation}

Therefore, the audio harmony can be formed as $h_a=\frac{h_s}{N'}$, while the visual harmony as $h_v=\frac{h_s}{M'}$, normalized by the numbers of music and motion beats. Inspired by the F-score \cite{sokolova2006beyond}, the final harmony score $h$ is obtained through a harmonic mean between $h_a$ and $h_s$, as:

\begin{equation}\label{finalequ}
h=\frac{(1+\beta^{2})h_{v}h_{a}}{\beta^{2}h_v+h_a}=\frac{(1+\beta^{2})\frac{{h_{s}}^{2}}{M'N'}}{\beta^{2}\frac{h_{s}}{M'}+\frac{h_{s}}{N'}} =\frac{(1+\beta^{2})h_{s}}{N'\beta^{2}+M'},
\end{equation}
where $\beta$ is a pre-defined constant controlling audio-visual balance. It can also be implied that, given $N'$ salient audio beats and $M'$ salient visual beats, the harmony score $h$ can be uniquely determined by $h_s$. 

As a heuristic evaluation strategy, we perform verification by introducing it as a regularizer during model training for the music-to-motion synthesis task. The validity of this harmony strategy can be confirmed if it successfully guides the motion generation process, producing high-quality cross-modal rhythmic consistency that aligns closely with subjective human perception.

\section{Learning harmony-aware human motion synthesis with music}

In this section, we propose a harmony-aware GAN model to efficiently tackle music-to-motion tasks under UGC training datasets. Our adversarial learning framework is built on rhythmic representations at both the meter and beat levels. Following network designs that employ an encoder-decoder-based generator with dual discriminators, we further introduce a 2D-to-3D pose lifting network to alleviate learning pressure. Additionally, our harmony evaluation strategy from Section \ref{AVH-Sec3} is incorporated into the model as a weakly-supervised loss component, which perceptually constrains cross-modal beat alignment during the synthesis process.

\subsection{Rhythm-oriented Adversarial Learning Framework}\label{fmw}

\begin{figure}[htbp]
\centerline{\includegraphics[scale=0.14]{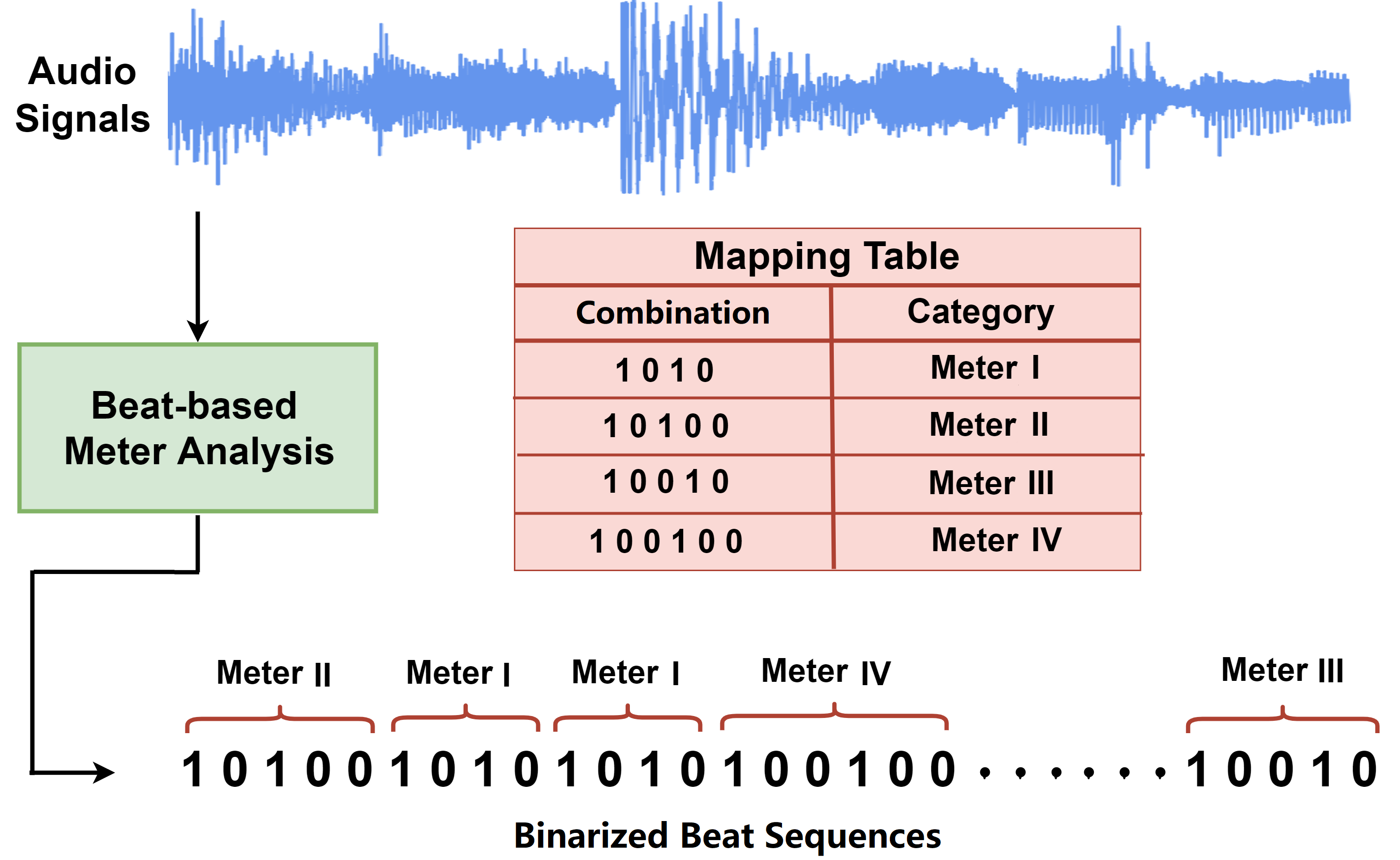}}
\caption{Mapping from audio to 4 types of meters based on beat analysis.}
\label{Metersep}
\end{figure}

\textbf{Meter segment pre-processing:} Previously, most deep music-to-motion methods focused primarily on low-level beat features \cite{tang2018dance,yalta2019weakly,lee2019dancing} as the rhythmic representations within their learning frameworks. However, the high-level connections inside the obtained beat sequence, known as the musical meter, are less addressed in the existing approaches. The concept of meter, composed of multiple strong and weak beats, effectively depicts melody by forming a periodic rhythmic unit \cite{benward2014music}. To enhance the model's high-level understanding of various musical rhythms, we segment the extracted audio beats into combinations and map them onto four types of commonly used meters.

As illustrated in Fig. \ref{Metersep}, given an audio sequence, we can derive the musical beats $B_a$ and their corresponding beat saliency $S_a$, as discussed in Section \ref{AVH-Sec3}. 
A beat is labeled $1$ as a strong beat if its saliency is greater than that of the preceding beat; otherwise, it is labeled $0$ to indicate a weak beat. Based on such an obtained binary beat sequence, we construct a mapping table that categorizes strong-weak beat combinations into 4 different meter types \cite{acuff1906sdn}. This allows us to flexibly handle music tracks containing either a single consistent meter or a mix of meters.

We preprocess the audio sequences by segmenting them into several meter units $MU=\{t \mid T_{start} \leq t \leq T_{end}\}$, each with a unified time length $T_u$. These segments include an entire musical meter and a few beats from the preceding meter to preserve inter-meter transitions. Our deep music-to-motion synthesis model is trained and tested on these segmented meter units.  We expect that by doing so, the model can implicitly capture high-level rhythmic features within the latent space, thereby enhancing generation robustness and efficiency even learned on small UGC training datasets.

\begin{figure*}[htbp]
\centerline{\includegraphics[scale=0.063]{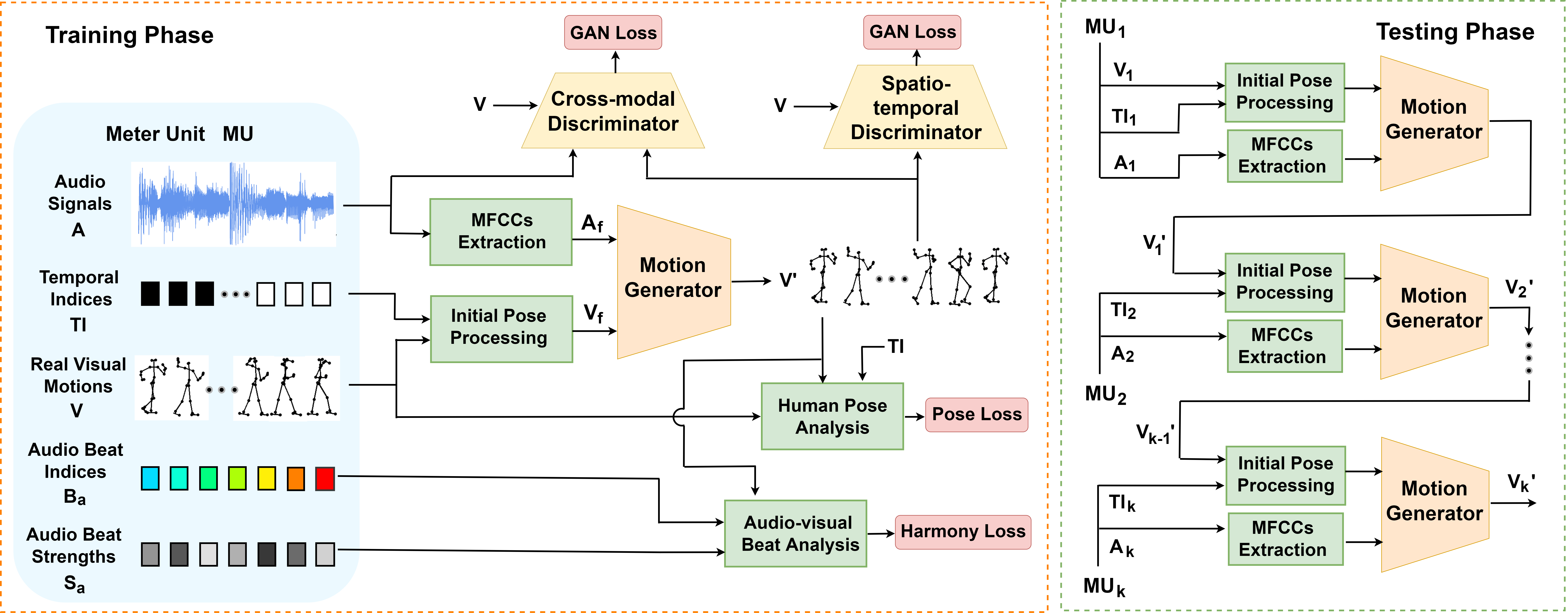}}
\caption{The overview of the whole framework, where the training and testing process are demonstrated in details.}
\label{OverView}
\end{figure*}
 
\vspace{0.5mm}
\textbf{Beat-driven adversarial framework:}
Given that GANs demonstrated their effectiveness for cross-modal tasks \cite{wang2019combining,sun2020deepdance,ferreira2021learning}, we leverage such an adversarial strategy in the learning framework. As shown on the left side of Fig. \ref{OverView}, our GAN model is trained using features processed from each meter unit  $MU$. During the time period of each $MU$, the audio sequence $A$ and ground-truth human motion pose $V$ are extracted from the UGC media, including the corresponding musical beats $B_{a}$ and their saliency $S_{a}$. 

In parallel, binary temporal indices $TI=\{e_{t} \mid T_{start} \leq t \leq T_{end}\}$ are collected to differentiate between initial poses and subsequent generated ones, where $e_{t}$ is set to $1$ if it fails within the preceding meter transitions. We derive the Mel Frequency Cepstral Coefficients (MFCCs) from $A$ as the auditory feature input $A_{f}$. Additionally, we obtain the visual feature input $V_{f}$ using $TI$ and $V$ through initial pose processing as:
\begin{equation}\label{Inipose}
 vf_{t}=
  \begin{cases}
    v_{t} & \text{if $e_{t}=1$} \\
    \frac{\sum_{t}{e_{t}v_{t}}}{\sum_{t}{e_{t}}} & \text{otherwise},
  \end{cases}
\quad \text{for } t = T_{start}, \dots, T_{end}
\end{equation}
where $v_{t}$ and $vf_{t}$ denote the visual representations at time $t$ for $V$ and $V_{f}$, respectively. This initialization $V_f$ provides motion priors to guide movement transitions, benefiting efficient cross-modal transformation with improved pose understanding across space and time.

In this way, our generator $G$ synthesizes motion sequences $V’$ by conducting $G(A_{f}, V_{f})=V’$. Two discriminators are employed to focus on cross-modal style and spatio-temporal consistency, which adversarially supervise the generation quality from different perspectives. Beyond the GAN losses provided by the discriminators, we incorporate a multi-space pose loss and a beat-based harmony loss to further regularize the motion realism and rhythmic synchronization during generation. These components will be detailed in the following subsections.

In the testing phase, as depicted on the right side of Fig. \ref{OverView}, our generator $G$ is able to recurrently synthesize human movements for an audio clip of arbitrary length segmented into $K$ meter units, using only a single initial pose input. This capability allows us to flexibly meet the demands of practical applications.

\subsection{GAN-based Network Architecture}

\begin{figure*}[htbp]
\centerline{\includegraphics[scale=0.07]{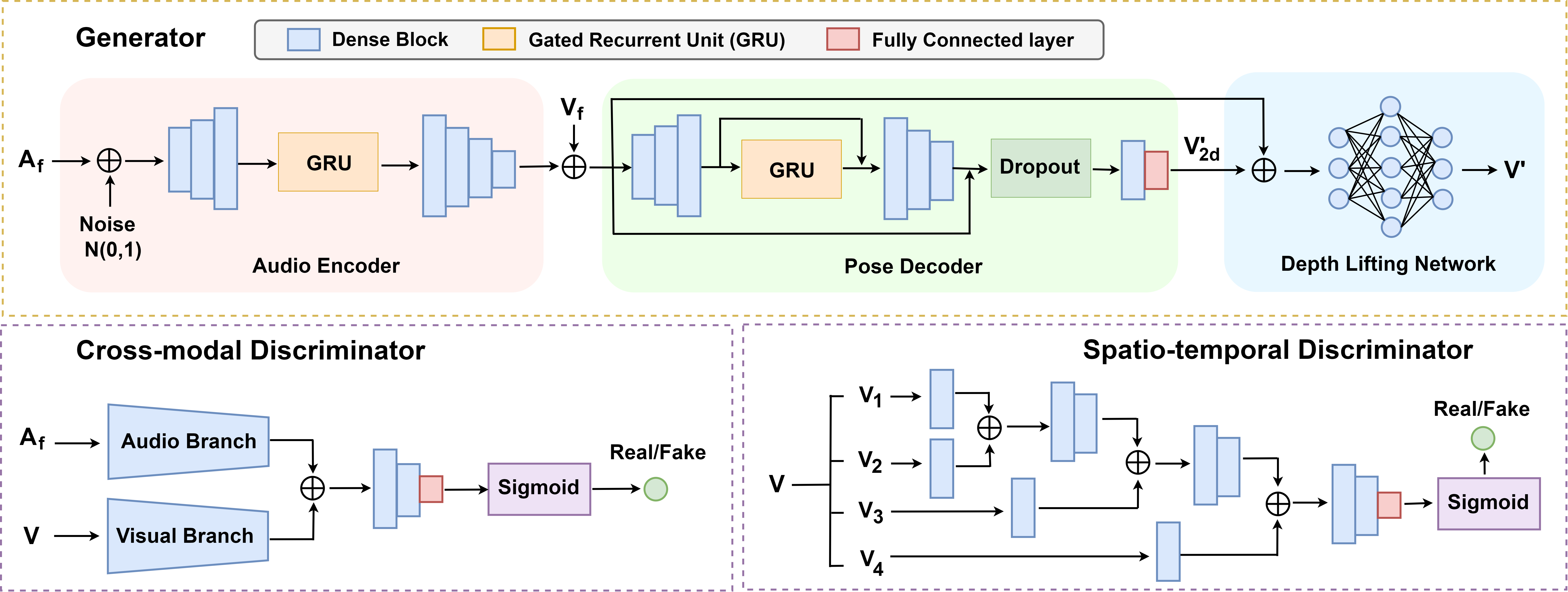}}
\caption{Detailed structural designs for all networks in our framework, where $\oplus$ denotes the concatenation operation.}
\label{NetArch}
\end{figure*}

\textbf{Generator:} As illustrated in the upper section of Fig \ref{NetArch}, our generator comprises an audio encoder, a pose decoder, and a depth lifting network so as to effectively manage feature translation between cross-modal sequences. Following existing designs \cite{lee2019dancing,ren2020self}, our audio encoder and pose decoder utilize dense blocks, each composed of a dense layer, layer normalization, and a LeakyReLu layer. We also leverage recurrent structures like Gated Recurrent Units (GRUs) to efficiently capture latent features with fewer parameters and reduced computation. In addition, we introduce random noise and dropout layers at the beginning and end of the encoding and decoding phases, respectively. This enhances model robustness, particularly under relatively fewer UGC training samples. 

Accurately estimating pose depth during motion generation is challenging. We thus additionally incorporate a depth lifting network to handle depth analysis, which builds on a 2D-to-3D lifting strategy widely used in image and video processing tasks \cite{tome2017lifting,Luo2020VideoDepth,kopf2020one}. Instead of directly synthesizing 3D human poses, our pose decoder first outputs preliminary 2D human poses $V’_{2d}$. These are then combined with previously obtained latent features to generate the final 3D poses $V’$. The high similarity between 2D and 3D poses enables our depth lifting network to focus on precise depth generation using 2D pose priors. We expect this strategy can enhance efficiency and reduce learning pressure of our generator.

\textbf{Cross-modal discriminator:} To ensure consistent content style between the generated human movements and the inputted audio sequences, our cross-modal discriminator, as shown on the bottom left of Fig. \ref{NetArch}, is designed using a two-branch classification network, inspired by previous methods \cite{lee2019dancing,ren2020self}. Each branch owns 3 sets of dense blocks to extract auditory and visual features independently. These features are subsequently concatenated together to estimate cross-modal stylistic consistency, which simultaneously benefits the high-level style understanding of our generator.

\textbf{Spatio-temporal discriminator:} Spatio-temporal consistency is crucial in video generation tasks and can be effectively captured by analyzing spatial features concatenated across different time spans \cite{lucas2019generative,chu2020learning}. Drawing inspiration from this, our spatio-temporal discriminator is created to constrain unrealistic motion generation, such as distorted spatial human poses or unnatural temporal transitions between movements. As shown on the bottom right of Fig. \ref{NetArch}, the input motion sequence $V$ is evenly segmented into four parts based on the time duration. Using a multi-level network, the temporally related movement features, extracted from each motion segment, are analyzed and combined in a local-to-global way. By progressively extracting spatial features and concatenating them temporally in chronological order, this discriminator judges the generation quality at both spatial and temporal levels, thereby implicitly guiding the generator to imitate the spatio-temporal pose flows in the ground truth data.

\subsection{Multi-domain Harmony-aware Loss Function}
To guide the generator in producing realistic human motions that are harmonious with the given music, our hybrid loss function consists of three components: pose loss, harmony loss, and GAN loss. These components provide supervision across multiple domains.

\vspace{0.5mm}
\textbf{Pose domain}: To ensure the realism of the generated human poses, we construct loss components that regularize the generator through the distribution, pixel, and feature spaces. 

For the distribution space, we use Kullback-Leibler (KL) divergence to supervise the accuracy of the generated preliminary 2D poses. The KL divergence between the probability distribution of the ground-truth 2D motion sequence $V_{2d}$ and our estimation $V’_{2d}$ is defined as:

\begin{equation}\label{kl}
\mathcal{L}_{kl}=KL(\mathbb{P}(V_{2d})||
\mathbb{P}(V_{2d}')),
\end{equation}
where $\mathbb{P}$ represents the probability distribution of the corresponding motion sequences.

In terms of the pixel space, an MSE loss component is employed to constrain the precision of 3D pose generation as:
\begin{equation}
\label{mse}
\mathcal{L}_{mse}=||(V-G(A_{f},V_{f}))||_{2},
\end{equation}
where the loss measures the difference between the ground-truth 3D poses $V$ and our generation $G(A_{f},V_{f})$.

To regularize the generated poses based on the feature space, we incorporate a VGG loss component. VGG networks \cite{simonyan2014very} are widely used to ensure that generated visual features align with human perception \cite{lucas2019generative,ma2021spatial}. We thus define the loss as:

\begin{equation}
\label{vgg}
\mathcal{L}_{feat}=||(VGG(V)-VGG(G(A_{f},V_{f})))||_{2}
\end{equation}
We assume this feature-space pose loss can capture the hidden kinematic correlations of motion flow so as to force our generation close to the ground truth.  

\vspace{0.5mm}
\textbf{Rhythm domain}: As discussed in Section \ref{AVH-Sec3}, the cross-modal rhythmic harmony between audio and human motion sequences can be measured using the evaluation strategy of audio-visual beat synchronization. To maximize audio-visual harmony in our generated results, we minimize the negative harmony score $-h$, which is derived based on Equations (\ref{basic}) and (\ref{finalequ}) as:

\begin{equation}
\begin{aligned}\label{lossharmo}
\mathcal{L}_{harmo}&=-h=-L(F_{a}(A),F_{v}(G(A_{f},V_{f}))) \\
& \approx -h_{s} + |M' - N'|,
\end{aligned}
\end{equation}
where all the variables have been defined in Section \ref{AVH-Sec3}. 

Since the harmony score $h$ heavily relies on $h_s$ (as referenced in Equation \ref{finalequ}), instead of directly applying Equation (\ref{finalequ}), here we leverage an efficient approximation $-h_{s}+|M'-N'|$ as our harmony loss component during training. It’s worthnoting that the term $|M'-N'|$ penalizes the over-frequent visual beats that may disrupt the audio-visual alignment. 
We expect this weakly-supervised loss component to not only enhance the rhythmic synchronization of the generated output but also improve the robustness of our model when trained on limited UGC samples.

\vspace{0.5mm}
\textbf{Adversarial domain}: Given that GANs adversarially supervise the generation by solving a min-max optimization problem during training, our cross-modal discriminator $D_{cd}$ and spatio-temporal discriminator $D_{st}$ try their best to distinguish between real ground-truth and synthesized sample pairs through maximizing their respective losses as:
\begin{equation}\label{lossdcd}
\begin{aligned}
{\mathcal{L}_{dcd}}&=\mathbb{E}[\log(1-D_{cd}(A_{f},G(A_{f},V_{f})))]\\
&+\mathbb{E}[\log D_{cd}(A_{f},V)], 
\end{aligned}
\end{equation}

\begin{equation}\label{lossdst}
{\mathcal{L}_{dst}} =\mathbb{E}[\log D_{st}(V)]+\mathbb{E}[\log(1-D_{st}(G(A_{f},V_{f})))].
\end{equation}

Conversely, our generator focuses on fooling the discriminators by minimizing the loss component as:
\begin{equation}\label{lossgan}
\begin{aligned}
{\mathcal{L}_{gan}}&=\mathbb{E}[-\log D_{cd}(A_{f},G(A_{f},V_{f}))]\\
&+\mathbb{E}[-\log D_{st}(G(A_{f},V_{f}))].
\end{aligned}
\end{equation}

To summarize, the overall loss function for the generator is constructed by combining all the individual loss components:
\begin{equation}
\begin{aligned}\label{losshbd}
\mathcal{L}_{total}&=\lambda_{kl}\mathcal{L}_{kl}+\lambda_{mse}\mathcal{L}_{mse}+\lambda_{feat}\mathcal{L}_{feat} \\
&+\lambda_{harmo}\mathcal{L}_{harmo}+\lambda_{gan}\mathcal{L}_{gan},
\end{aligned}
\end{equation}
where each $\lambda$ denotes the weight balancing the corresponding loss component.

By utilizing such a multi-domain harmony-aware loss function, we assume the generator can synthesize high-quality human motions from various aspects, where the performance of our model will be validated in the next section.

\section{Experiments and results}
In this section, we first describe the experimental details of our proposed HarmoGAN. The generated motion sequences are then evaluated based on two datasets compared to state-of-the-art models in terms of motion realism, style consistency and rhythmic harmony both quantitatively and qualitatively. 

\subsection{Dataset}
We train our HarmoGAN using the Dance dataset, a relatively small collection created by a professional dancer, as introduced in \cite{tang2018dance}. This dataset allows us to simulate customized UGC data for practical scenarios. It includes 94 minutes of dance video, totaling 141,000 frames at 25 fps, along with 21-joint 3D human body keypoints and the corresponding audio tracks. There are 4 typical types of dances in the dataset: cha-cha, rumba, tango, and waltz. To enhance training efficiency, all videos are resampled at 15 fps to form training pairs without significantly compromising perceptual quality. Approximately 75 minutes of video are used for training, while the remaining 19 minutes are reserved for evaluation.

During the testing phase, we additionally utilize the public Ballroom music dataset \cite{gouyon2006experimental} to further assess the effectiveness of our model in enhancing cross-modal rhythmic harmony. It contains 698 background music clips extracted from online dance videos, involving 7 types: cha-cha, jive, quickstep, rumba, samba, tango, and waltz. Due to the repetitive nature of musical melodies, we randomly select 10\% of audio sequences from each dance category, which are cut into 6-second clips, to form our testing dataset.

\subsection{Implementation Details}
All the networks and loss functions are implemented in PyTorch. The generator of HarmoGAN is first pretrained with 225 epochs using $\mathcal{L}_{pretrain}=0.1\mathcal{L}_{kl}+\mathcal{L}_{mse}$. The Adam optimizer \cite{kingma2014adam} is applied with a batch size of 10. The initial learning rate is set to 0.001 and then gradually decreases after every 50 epochs, following a decay factor schedule [0.5,0.2,0.2,0.5]. Initializing with the pretrained model, the GAN training involves both generator and discriminator networks. The weights of losses are set as follows: $\lambda_{kl}=0.0001, \lambda_{mse}=\lambda_{feat}=\lambda_{gan}=0.001,$ and $\lambda_{harmo}=1$. The learning rates for all networks are initialized at 0.0001 and divided by 2 and 5 alternatively after every 5 epochs. The weight decay is set to 0.001 for the discriminators and 0.0001 for the generator. The optimizer and batch size are kept the same as in the pretraining. After 45 epochs of adversarial training, our HarmoGAN converged successfully.

For our proposed harmony evaluation strategy, the constant factors $\lambda_1$ and $\lambda_2$ in Equations (\ref{filta}) and (\ref{filtv}), respectively, are set to 0.1 and 1 to achieve the optimal cross-media beat alignment for the ground truth. The reaction delay $T_{delay}$ in Equation (\ref{Tdelay}) is set to 0.25 seconds following findings in \cite{ho2013extraction} to simulate human visual perception precisely. $\beta$ in Equation (\ref{finalequ}) is set equal to 2 for maintaining an appropriate audio-visual balance.

\begin{figure*}[htbp]
\centerline{\includegraphics[scale=0.07]{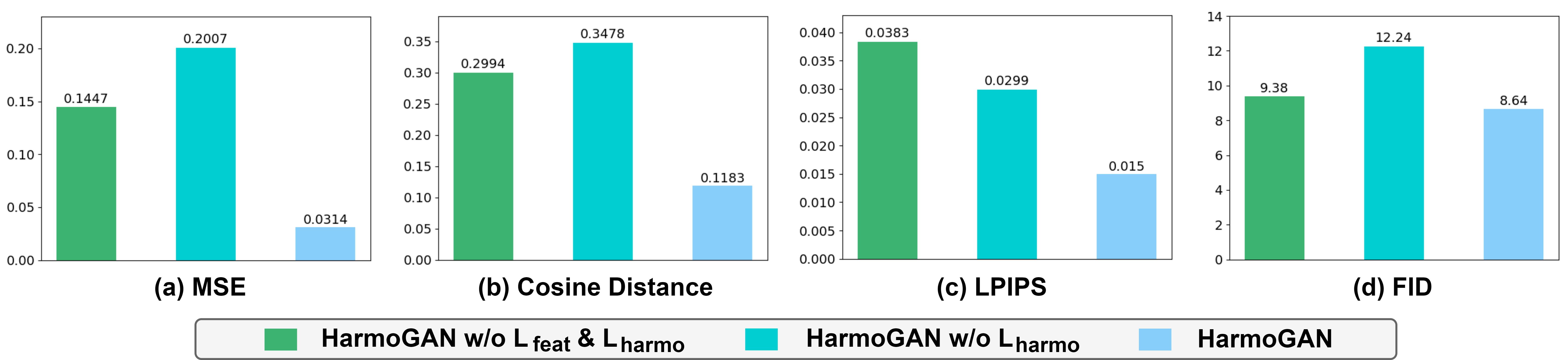}}
\caption{Comparison of motion quality between HarmoGAN and its other GAN variants based on multiple quantitative metrics. All the metrics are the lower the value the better.}
\label{abl_mq}
\end{figure*}

\subsection{Evaluation of Generated Motion}

To assess the quality of our generated human motion poses, we test our HarmoGAN on the Dance dataset \cite{tang2018dance} using ground-truth dance movements as a reference. Our model is compared against two other GAN variants, each getting rid of certain loss components during training. We utilize metrics including MSE, Cosine Distance, LPIPS \cite{simonyan2014very} and FID \cite{heusel2017gans} to measure the accuracy of the generated motions. The first two metrics focus on assessing the spatio-temporal coherence of the poses, while the latter two evaluate the style consistency of the cross-modal transformation through deep feature analysis.

As illustrated in Fig. \ref{abl_mq},  when the training process is dominated solely by $\mathcal{L}_{gan}$, the variant model (labeled in green) shows some improvements in MSE, Cosine Distance and FID but significantly fails in LPIPS. On the other hand, incorporating $\mathcal{L}_{feat}$ (indicated by the model in cyan) benefits LPIPS but degrades in other metrics. Meanwhile, a relatively small size of our training dataset can lead to issues such as overfitting and missing the global minimum. The introduction of $\mathcal{L}_{harmo}$ (represented by the model in light blue), which is a wealy-supervised loss component, yet effectively guides our HarmoGAN towards the best trading-off between all the metrics, resulting in the high-quality motion generation.    

\begin{figure}[htbp]
    \centering
   \includegraphics[scale=.09]{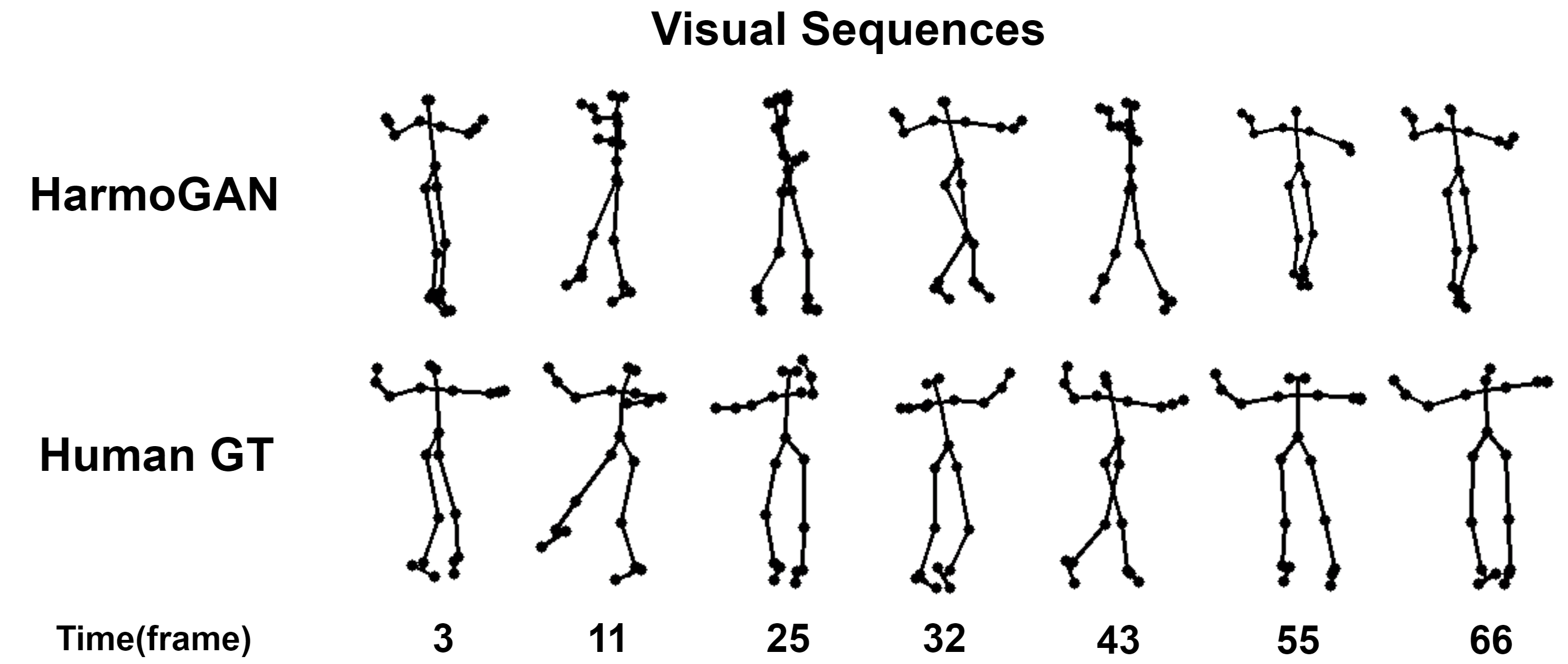}
    \caption{Qualitative comparison based on the Dance dataset to demonstrate the natural pose flows of our generation.}
    \label{realism-quali}
\end{figure}

In Fig. \ref{realism-quali}, we also provide a qualitative example comparing our generation with the corresponding human ground truth. It can be observed that results synthesized by our HarmoGAN present realistic body movements and consistent styles that resemble those from the human dancer.

\subsection{Evaluation of Harmony Assessment Strategy}\label{harmopercep}

We evaluated the validity of our harmony assessment strategy proposed in Section \ref{AVH-Sec3} through a user study. To ensure that this strategy accurately measures rhythmic harmony in a way consistent with human perception, we collected 20 groups of side-by-side videos for qualitative analysis. Each group contains two video clips: one sampled from the ground truth of the Dance dataset and another created by intentionally permuting or warping the original visual sequences. Twelve participants were invited to watch all 20 video pairs and provide their perceptual judgments by selecting the clip they considered more harmonious in each pair. These selections were then compared with the scores generated by our harmony assessment strategy. 

\begin{figure}[htbp]
    \centering
   \includegraphics[scale=.068]{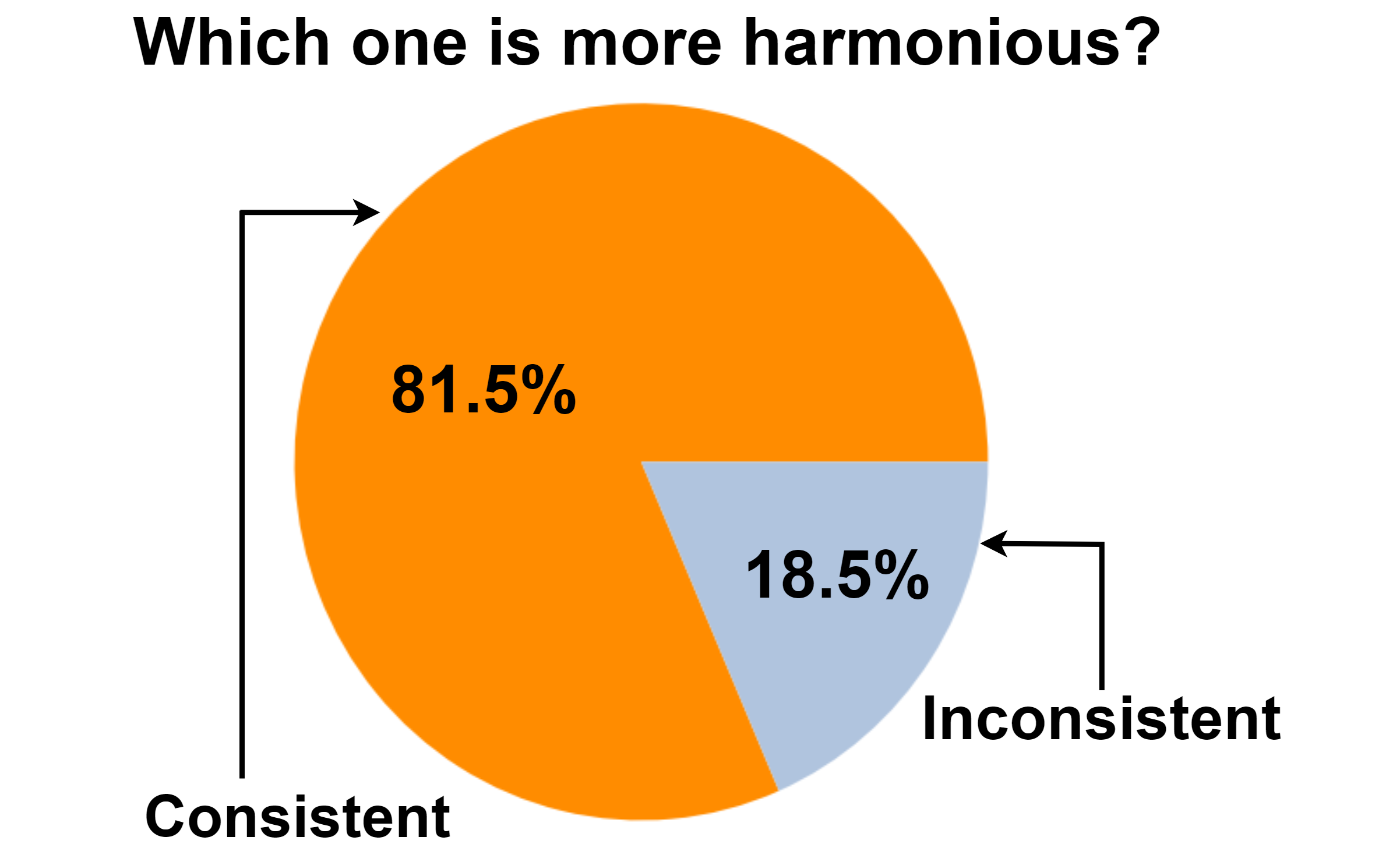}
    \caption{Percentage of consistency between human judgments and our proposed harmony assessment strategy for evaluating rhythmic harmony.}
    \label{harmorecog}
\end{figure}

Fig. \ref{harmorecog} presents the results showing the average consistency percentage across all participants and test groups. It can be seen that our harmony strategy aligned with human judgments in 81.5\% of the cases. The remaining inconsistencies may be attributed to factors such as biological limitations, personal preferences, or a lack of background knowledge among some participants, leading to insensitivity to rhythmic harmony. Overall, these results provide reasonable validation for the effectiveness of our harmony assessment strategy. Additionally, we provide live user study samples in the supplementary video, demonstrating how the harmony score precisely captures unsynchronized cross-modal rhythms.

\subsection{Evaluation of Rhythmic Harmony}

In the following subsections, we assess the improved performance of rhythmic harmony in our HarmoGAN, particularly focusing on the impact of the incorporated harmony loss $\mathcal{L}_{harmo}$, through both quantitative and qualitative comparisons with variant models and state-of-the-art approaches.

\vspace{0.5mm}
\textbf{Quantitative analysis on the Dance dataset:} The rhythmic harmony of our generated results is evaluated using two metrics: the harmony score $h$ (see Equation (\ref{finalequ}) for details) and the hit rate \cite{lee2019dancing,sun2020deepdance}. Following the human perceptual principle, the hit rate here counts the percentage of audio beats that are synchronized with visual beats within the duration of the perception delay. We compare the generation of our HarmoGAN with its variant model without $\mathcal{L}_{harmo}$ and the ground-truth motion sequences from the Dance dataset performed by a human dancer.

\begin{table*}[htbp]
\centering
\caption{Harmony score for 7 types of dance music}
\label{quantiharmo}
\begin{tabular}{c c c c c c c c|c}
\hline
Model & Cha-cha & Jive & Quickstep & Rumba & Samba & Tango & Waltz & Average \\
\hline
Lee et al. \cite{lee2019dancing} & 0.3509 & 0.3359	& 0.2773 & 0.2862 & 0.2805 & 0.2657 & 0.2703 & 0.2953 \\
Ren et al. \cite{ren2020self} & 0.2759 & 0.3321 & 0.1625 & 0.2154 & 0.2761 & 0.2671 & 0.3122 & 0.2630 \\
HarmoGAN w/o $\mathcal{L}_{harmo}$ & 0.1536 & 0.2115 & 0.2150 & 0.1679 & 0.2495 & 0.1929 & 0.2377 & 0.2040 \\
HarmoGAN & \textbf{0.4097} & \textbf{0.3995} & \textbf{0.3199} & \textbf{0.3529} & \textbf{0.3468} & \textbf{0.2948} & \textbf{0.3455} & \textbf{0.3527} \\
\end{tabular}
\end{table*}

\begin{table*}[htbp]
\centering
\caption{Hit rate for 7 types of dance music}
\label{hitrate-ab}
\begin{tabular}{c c c c c c c c|c}
\hline
Model & Cha-cha & Jive & Quickstep & Rumba & Samba & Tango & Waltz & Average \\
\hline
Lee et al. \cite{lee2019dancing} & 58.89\%	& 59.92\% & 53.03\% & 65.06\% & 64.92\% & 58.46\% & 54.30\%  & 59.23\% \\
Ren et al. \cite{ren2020self} & 28.13\% & 51.03\% & 33.33\% & 41.53\% & 49.07\% & 42.84\% & 52.59\% & 42.65\% \\
HarmoGAN w/o $\mathcal{L}_{harmo}$ & 14.25\% & 17.45\% & 18.18\% & 20.91\% & 29.12\% & 21.48\% & 25.53\% & 20.99\% \\
HarmoGAN & \textbf{74.60\%} & \textbf{75.70\%} & \textbf{71.21\%} & \textbf{77.01\%} & \textbf{83.69\%} & \textbf{74.98\%} & \textbf{72.89\%} & \textbf{75.73\%} \\
\end{tabular}
\end{table*}

\begin{figure}[htbp]
    \centering
   \includegraphics[width=0.98\linewidth]{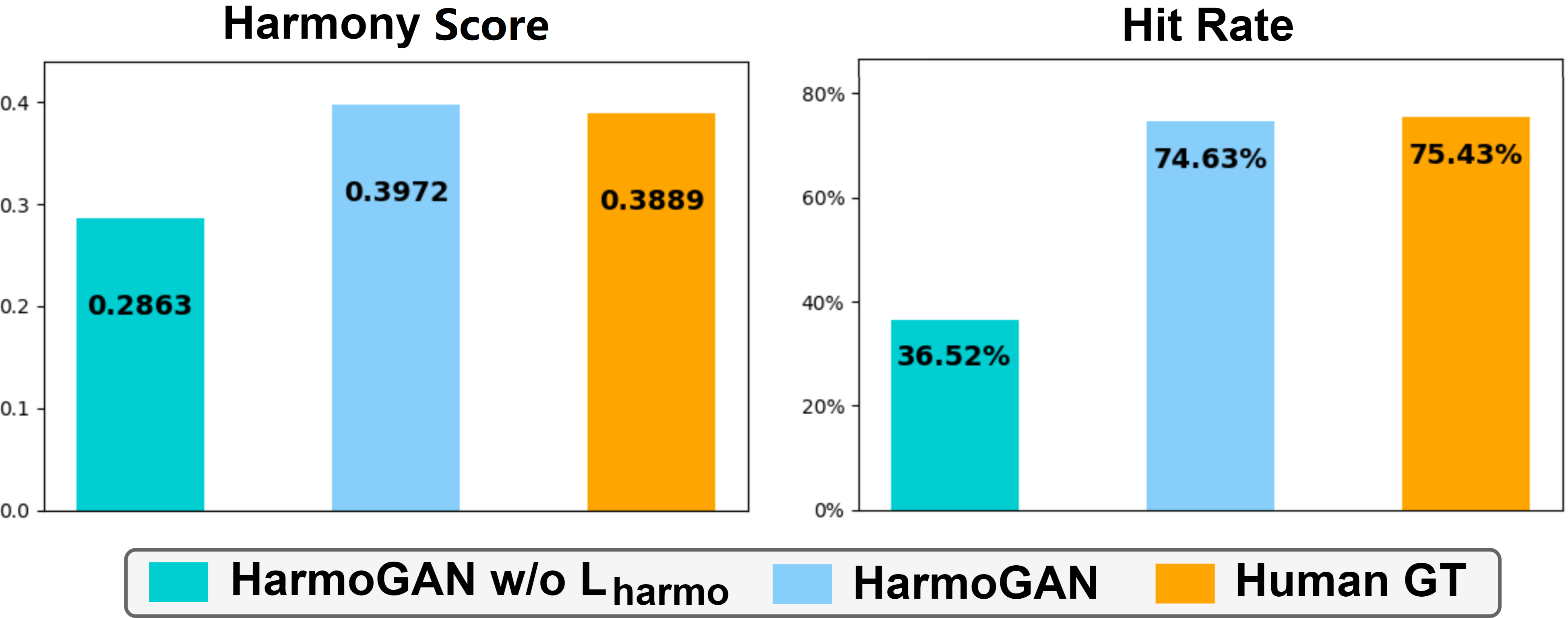}
    \caption{Comparison of cross-modal rhythmic harmony on the Dance dataset using harmony score and hit rate of audio beats. In both cases, a higher metric value indicates a better result.}
\label{abl2}
\end{figure}

As shown in Fig. \ref{abl2}, the testing results demonstrate that HarmoGAN, with the incorporation of $\mathcal{L}_{harmo}$, significantly outperforms its variant in both harmony score and hit rate, achieving results that are comparable to human ground truth. Due to the constrained accuracy of personal data acquisition devices, which is common in UGC video production, even human dancers may struggle to achieve optimal performance in quantitative evaluations, as noted in previous studies \cite{lee2019dancing}. Yet, the weakly-supervised manner of our $\mathcal{L}_{harmo}$ effectively alleviates this problem. By guiding the model through the audio-visual beat alignment strategy, $\mathcal{L}_{harmo}$ flexibly and robustly leads to harmonious motion synthesis rather than simply forcing the model to replicate the limited and noisy ground truth.

\begin{figure}[htbp]
    \centering
   \includegraphics[scale=.12]{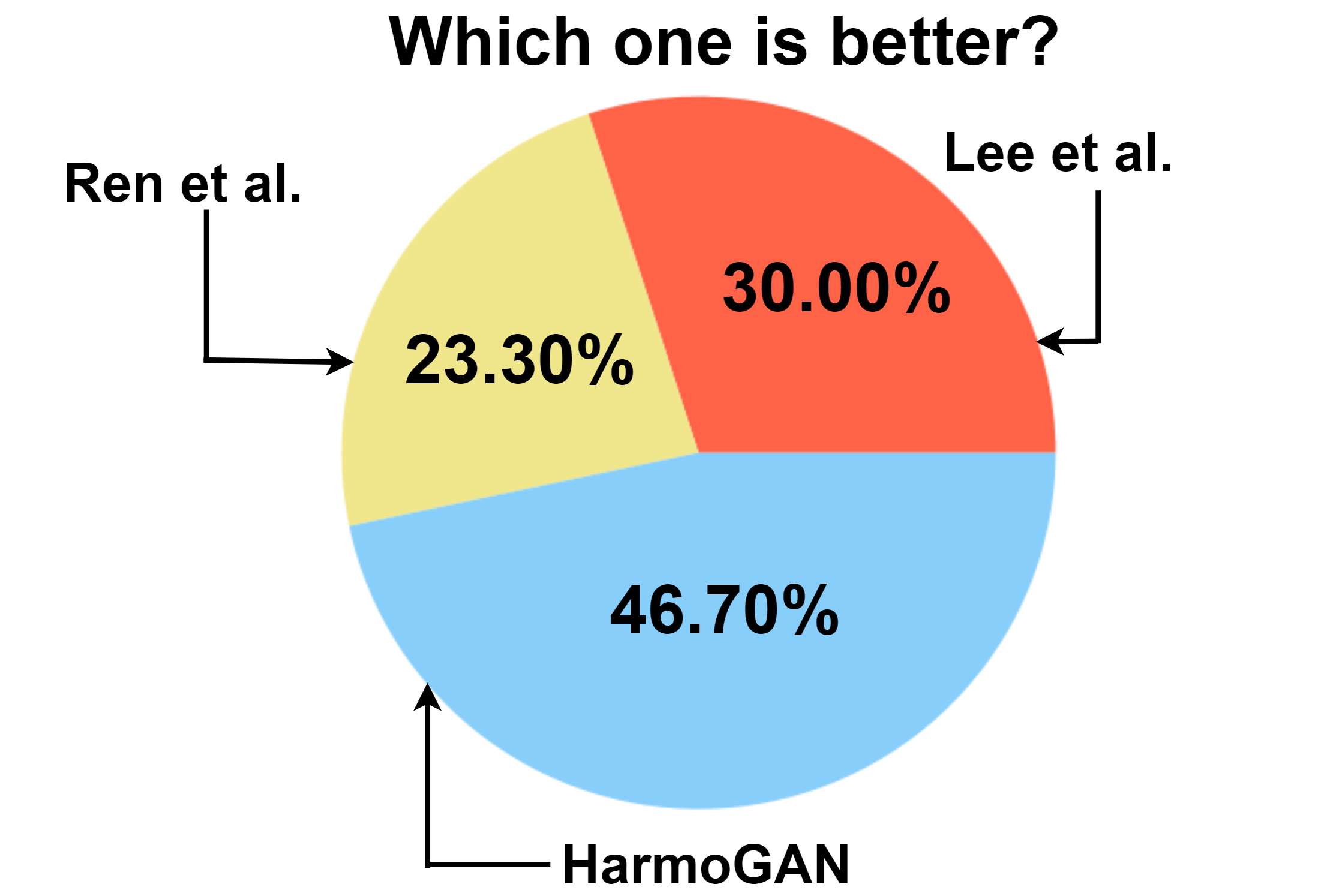}
    \caption{Average preference percentages across 10 groups of side-by-side video comparisons.}
\label{abtest}
\end{figure}

\begin{figure*}[htbp]
\centerline{\includegraphics[scale=0.17]{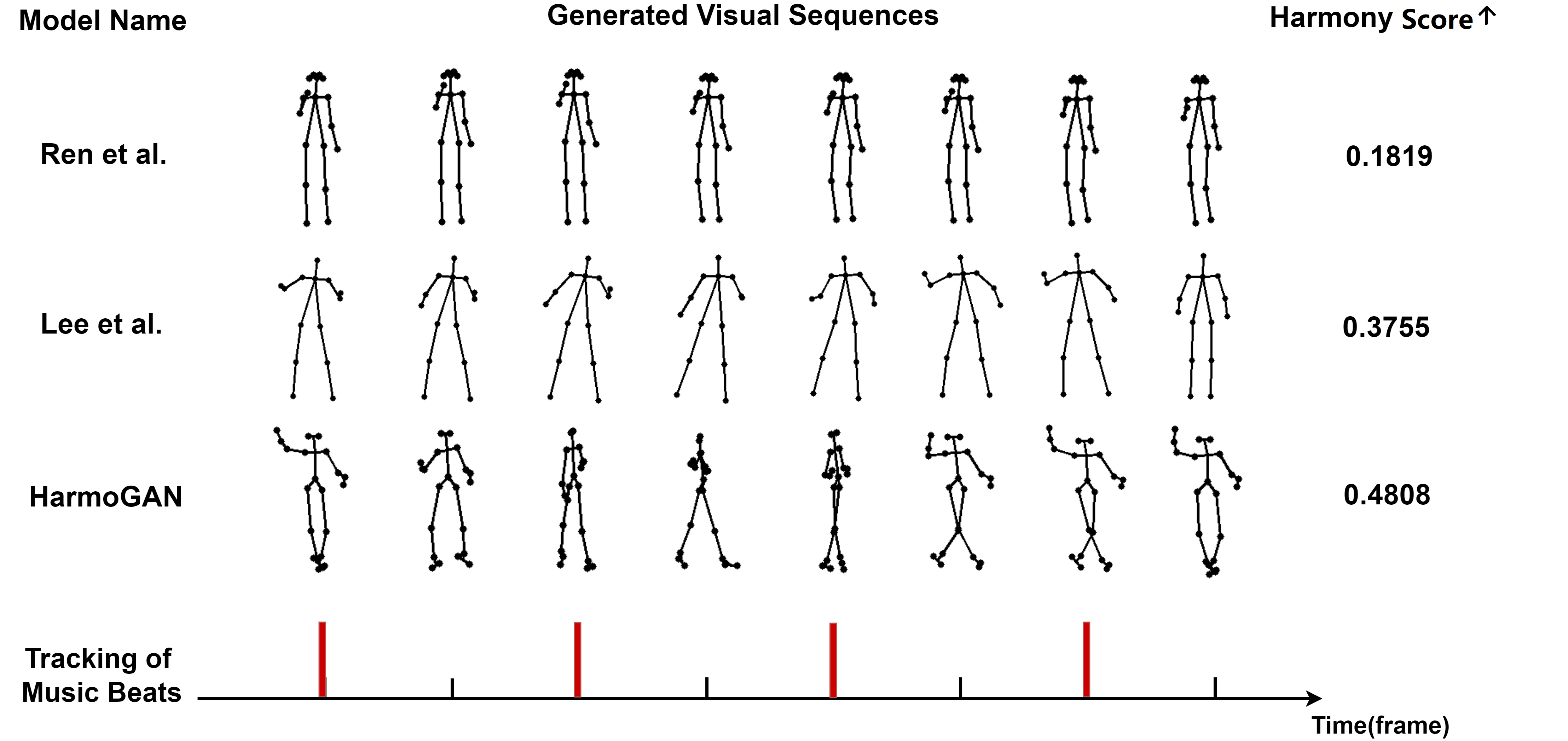}}
\caption{Qualitative comparison example of generated motion sequences with corresponding audio beats (highlighted in red) to illustrate cross-modal rhythmic harmony achieved by different models.}
\label{qualitest}
\end{figure*}

\vspace{0.3mm}
\textbf{Quantitative comparison on the Ballroom dataset:} To further evaluate the effectiveness of HarmoGAN in controlling audio-visual rhythmic harmony, we compare our model with two leading music-to-motion synthesis approaches proposed by Lee et al. \cite{lee2019dancing} and Ren et al. \cite{ren2020self}, using their published codes and models. For a fair comparison to discuss the effects of $\mathcal{L}_{harmo}$, all tested models are GAN-based and assessed on the Ballroom music dataset \cite{gouyon2006experimental}, which is not included in the training phase of each method.

We synthesize motion sequences based on music tracks from the Ballroom dataset and combine the cross-modal media to evaluate rhythmic harmony using metrics including harmony score and hit rate. The results for each metric are shown in Tables \ref{quantiharmo} and \ref{hitrate-ab}, covering seven types of dance music. Our HarmoGAN outperforms the other models across all music types and metrics. Notably, without the guidance provided by the weakly-supervised $\mathcal{L}_{harmo}$, the variant model significantly degrades when tested on an untrained dataset.

Compared to our variant model, the implicit cross-modal synchronization learned through the self-attention discriminator in Ren et al. \cite{ren2020self} offers some improvement in rhythmic harmony. However, these improvements are closely tied to the dance types present in the training set and are less effective for specific types like “Cha-Cha” and “Quickstep,” as shown in the tables. Lee et al. \cite{lee2019dancing} generally outperform Ren et al. \cite{ren2020self} in both metrics by applying a post-processing beat warper. However, this approach can hardly adjust generated motions in real-time during the synthesis process, leading to limited enhancements in harmony scores, likely due to a lack of synchronization with salient audio beats.

In contrast, the utilization of a high-level beat alignment strategy in our HarmoGAN enables us to efficiently control rhythmic harmony on the way of model generation and provide stable improvements regardless of the diversity in dance types.   

\vspace{0.5mm}
\textbf{User study:}\label{qualistudy} We also conduct a user study to qualitatively assess the overall quality of our generated motion sequences in comparison with those produced by Lee et al. \cite{lee2019dancing} and Ren et al. \cite{ren2020self}. Twelve participants are invited to join the study. Each of them is asked to watch 10 groups of dance video pairs, which are synthesized from 3 different models using the same music, and then blindly select the best video clip considering both motion realism and rhythmic harmony.

As shown in Fig. \ref{abtest}, our HarmoGAN receives more preferences compared to the other two models, despite individual differences in aesthetics and sensitivity. This implies that cross-modal rhythmic harmony plays an important role in human perception, significantly influencing the overall quality judgment of dance videos, in particular under similar motion generation capabilities. These results further validate the effectiveness of our beat-based harmony evaluation strategy and the incorporated harmony loss $\mathcal{L}_{harmo}$.

A qualitative example is provided in Fig. \ref{qualitest} using a music sample from the Ballroom music dataset. In this example, the motions generated by Ren et al. \cite{ren2020self} exhibit perceptually indistinct visual beats in the body movements. In comparison, the generation by Lee et al. \cite{lee2019dancing} shows more observable visual beats, yet some movement changes are still not salient enough to form valid visual beats that synchronize with the corresponding audio beats. Conversely, HarmoGAN produces distinct pose changes during motion synthesis, creating attentional visual beats that closely align with the audio beats, resulting in a higher harmony score. Additional live comparison results can be found in the supplementary video, which demonstrates the realism, diversity, and audio-visual harmony of our synthesized motion sequences for qualitative evaluation.

\section{Discussion}

HarmoGAN has demonstrated exceptional capability in achieving rhythmic harmony during music-to-motion synthesis, particularly in terms of practical scenarios. Our method has shown that high-quality cross-modal generation is possible even with a small UGC dataset. Moving forward, we plan to explore the potential enhancement of capabilities by adapting our weakly-supervised harmony strategy to larger datasets and more advanced networks, such as transformers. This will allow us to investigate the optimal trade-off between increased computational resources and the effectiveness of audio-to-visual synchronization. Although our experimental results primarily focus on dance sequences, the beat-based rhythm alignment approach we propose has broader applicability. It can be extended to other audio-visual tasks requiring cross-modal rhythmic consistency, such as speech-to-lip synchronization or music-driven video synthesis, which are also promising future directions.

\section{Conclusion}

In this paper, we propose a harmony-aware GAN framework for high-quality music-driven motion synthesis in UGC scenarios, aiming to enhance rhythmic harmony and improve viewer engagement. Drawing on principles of human perception, we analyze beat-based rhythmic representations and design a novel interval-driven cross-modal beat alignment strategy to capture rhythmic harmony. To cooperate with such a harmony strategy, we developed an encoder-decoder GAN architecture, named HarmoGAN, which incorporates dual discriminators and a depth-lifting subnetwork to effectively generate natural 3D human motion. Our HarmoGAN is subsequently trained in an adversarial manner based on categorized musical meter segments using a UGC training set. Additionally, we incorporate the proposed harmony strategy as a weakly-supervised perceptual constraint to align the generated audio-visual beat pairs for improving audio-visual harmony. This enables flexible synchronization of cross-modal rhythms during the synthesis process within a high-level rhythmic space covering both beats and meters. Experimental results demonstrate that our HarmoGAN can produce realistic human movement sequences even with a relatively small training dataset. Moreover, our model significantly outperforms other leading music-to-motion models in both quantitative and qualitative evaluations regarding audio-visual rhythmic harmony.


%

\bibliographystyle{IEEEtran}
\bibliography{biblio}

\end{document}